\newcommand{\braket}[1]{\left<#1\right>}

\newcommand{\cpp}{C\nolinebreak\hspace{-.05em}\raisebox{.4ex}\,\nolinebreak\hspace{-.10em}\raisebox{.25ex}{\footnotesize\bf
		++}\,}
\def\cpp{{C\nolinebreak[4]\hspace{-.05em}\raisebox{.25ex}{\footnotesize\bf
				++}}\,}

\newcommand{\dd}{\mathrm{d}}

\newcommand{\rnp}{r_\mathrm{np}}

\newcommand{\eV}{\mathrm{eV}}
\newcommand{\keV}{\mathrm{keV}}

\newcommand{\nm}{\mathrm{nm}}
\newcommand{\um}{\mu\mathrm{m}}

\newcommand{\cm}{\mathrm{cm}}

\newcommand{\Gy}{\mathrm{Gy}}
\newcommand{\mGy}{\mathrm{mGy}}

\newcommand{\smalltimes}{\scriptsize{\times}}
\def\cpp{{C\nolinebreak[4]\hspace{-.05em}\raisebox{.4ex}{\tiny\bf ++}}\ }

\documentclass[10pt, a4paper, unknownkeysallowed]{extarticle}
\usepackage[left=2cm,right=2cm,top=2cm,bottom=2cm,includeheadfoot]{geometry}

\usepackage{amssymb}			
\usepackage{amsmath}
\usepackage[makeroom]{cancel}

\usepackage{xcolor}

\usepackage{mathtools}
\usepackage{multicol}
\usepackage{multirow}
\usepackage{float}
\usepackage{hyperref}
\usepackage{cite}
\usepackage{comment}

\usepackage[most]{tcolorbox}
\usepackage{graphicx}
\usepackage[export]{adjustbox}

\usepackage[font=small,labelfont=bf, figurename=Fig.]{caption}
\usepackage[subrefformat=parens,labelformat=parens]{subfig}

\DeclareCaptionLabelFormat{mainfigure}{\figurename~#2}
\DeclareCaptionLabelFormat{maintable}{\tablename~#2}
\DeclareCaptionLabelFormat{supplementaryfigure}{Supplementary~\figurename~#2}
\DeclareCaptionLabelFormat{supplementarytable}{Supplementary~\tablename~#2}
\usepackage{natbib}

\begin{document}
\begin{center}	
	\begin{minipage}{13cm}
		\begin{center}	
			\setlength\parindent{5pt}
			\newcommand{\thetitle}{Radial dependence of ionization
				clustering around a gold nanoparticle}
			\vspace{.5cm}
			\textbf{\Large{\thetitle}}\\
			\vspace{.5cm}
			\small Leo Thomas$^\ast$,
			\small Miriam Schwarze,
			\small Hans Rabus
			\small \\
			\textit{Physikalisch-Technische Bundesanstalt (PTB),
				Abbestr. 2-12,
				D-10587 Berlin, Germany} \\
			\vspace{2mm}
			\hrule
			\vspace{2mm}
            $^\ast$ E-mail: \href{mailto:leo.thomas@ptb.de}{leo.thomas@ptb.de}
			\vspace{.5cm}
		\end{center}
	\end{minipage}
\end{center}

\setcounter{section}{0}

\begin{center}
	\centering
	\begin{minipage}{13cm}

\begin{center}
	\textbf{Abstract}
\end{center}

\textit{Objective:}
This work explores the enhancement of ionization clusters around a gold nanoparticle
(NP), indicative of the induction of DNA lesions, a potential trigger for
cell-death.

\textit{Approach:}
Monte Carlo track structure simulations were performed to determine (a) the fluence
of incident photons and electrons in water around a gold NP under charged particle
equilibrium conditions and (b) the density of ionization clusters produced on
average as well as conditional on the occurrence of at least one interaction in the
nanoparticle using Associated Volume Clustering. Absorbed dose was determined for
comparison with a recent benchmark intercomparison. Reported quantities are
normalized to primary fluence, allowing to establish a connection to macroscopic
dosimetric quantities.

\textit{Main results:}
The modification of the electron fluence spectrum by the gold NP is minor and mainly
occurs at low energies. The net fluence of electrons emitted from the NP is
dominated by electrons resulting from photon interactions. Similar to dose
enhancement, increased ionization clustering is limited to a distance from the NP
surface of up to $200\,\nm$. Smaller NPs cause noticeable peaks in the conditional
frequency of clusters at distances around $50\,\nm$ to $100\,\nm$ from the NP
surface. The number of clusters per energy imparted is increased at distances of up
to $150\,\nm$, and accordingly the enhancement in clustering notably surpasses that
of dose enhancement.

\textit{Significance:}
This work highlights the necessity of nanodosimetric analysis and suggests increased
ionization clustering near the nanoparticles due to the emission of low energy Auger
electrons. Whereas the electron component of the radiation field plays an important
role in determining the background contribution to ionization clustering and energy
imparted, the dosimetric effects of nanoparticles are governed by the interplay of
secondary electron production by photon interaction (including low energy Auger
electrons) and their ability to leave the nanoparticle.

	\end{minipage}
\end{center}

\vspace{0.5cm}

\section{Introduction}
\label{section:introduction}

In radiation therapy of tumors, well-established strategies to expand the
therapeutic window are to optimize tumor conformity to the tumor volume and increase
the tumor\textquotesingle s susceptibility to irradiation. This can be done by
either selectively enhancing the absorbed dose or increasing the tumor
cells\textquotesingle\ sensitivity to irradiation, respectively. The introduction of
gold nanoparticles (NPs) into the targeted region has been proposed as a means to
achieve in both goals and has been explored in both in-vitro and in-vivo trials on
mice \citep{Hainfeld_2013}.

The augmentation effect is particularly manifest in X-ray-based treatment where the
presence of gold NPs leads to a two-fold effect: an increase of the average dose to
the tumor volume for the same fluence of incident photons and a locally increased
dose in the proximity of the gold NP at distances of up to $200\,\nm$ from the
nanoparticle surface \citep*{McMahon_2011, Rabus_2019, Rabus_Li_2021}. While the
average dose enhancement is predominantly induced by the generation of photo- and
Compton electrons within the gold NP, the proximal effect is attributed to
de-excitation cascades, following the emission of a photoelectron, which give rise
to Auger(-Meitner) electrons with energies below $15\,\keV$ (L and M shells) (K
shell vacancies are predominantly filled by fluorescence transitions creating L
shell vacancies \citep{EADL}.)

The spatial distribution of energy imparted in the vicinity of an interacting
nanoparticle is determined by multiple factors such as nanoparticle size
\citep{Zygmanski_2013,Koger_Kirkby_2016, McMahon_2011}, coating
\citep{Belousov_2019, Morozov_2018} and the fluence of incident photons. The
fluence-dependence is two-fold: the probability of a photon interaction in a gold NP
is proportional to the total fluence \citep{Zygmanski_2013} and the proportionality
factor depends qualitatively on the energy distribution of incident photons
\citep{Kolyvanova_2021}. Notably, the proximal and peripheral effect depend on these
factors distinctively, cf. the `dichotomous nature' of nanoparticle size described
in \cite{Gadoue_2018}.

With respect to the proximal effect, previous studies have focused on the spatial
distribution of energy imparted (or related quantities as the local dose enhancement
factor) \citep{Mesbahi_2010,Zygmanski_2016,Moradi_2020, Vlastou_2020, Rabus_Li_2021}
or dose spikes \citep{Lin_2015b, Poignant_2021}. However, DNA lesion clusters have
been identified as a driving factor of cell-death inducing damage
\citep{Lomax_2013}. While DNA lesion clustering has gathered more attention in the
context of high-LET radiation (cf. \citep{Rucinski_2021}), low-energy electrons such
as Auger-Meitner electrons may similarly lead to a local increase in ionization
clusters.


While DNA lesions can be studied at the cellular level with immunofluorescence
assays \citep{Gonon_2019} and electron microscopy \citep{Lorat_2015}, measuring
ionization clustering, directly as a physical phenomenon, is restricted to
experiments simulating nanometric volumes exploiting density scaling principles
\citep{Bantsar_2018}. Investigations using Monte Carlo simulations to realistically
model radiation interactions around a gold NP with feasible efficiency are
challenging and require the application of variance reduction techniques.

Such simulations have been performed in a multi-step paradigm, where in a first
simulation the fluence at the nanoparticle is determined in water, after which
radiation transport within the gold nanoparticle and the ensuing effects in the
water surrounding it are simulated (in one or two steps) \citep{Lin_2014,
	Velten_Tome_2023}. Typically, phase-space files are employed to interface the
different simulations \citep{Klapproth_2021}.

While also employing a two-step approach, this work differs from previous studies in
several aspects. First of all, a nanodosimetric perspective is taken by
investigating the spatial distribution of ionization clusters (cf.
Section~\ref{subsection:nanodosimetry}), rather than energy imparted alone. Second,
in the preliminary simulation step, energy distributions of photons and electrons
are determined by scoring step lengths in a region of interest to obtain the
spectral fluences of the two particle types without the need for manipulating
phase-space files. Third, track-structure simulation is used for electron transport
and the simulation setup is such that charged particle equilibrium (CPE) conditions
are fulfilled. A lack of CPE could otherwise lead to significantly biased results
\citep{Rabus_2019, Rabus_2021, Rabus_2024}. Beyond previous efforts, the photon- and
electron fluences impinging on the nanoparticle are studied individually and for
different nanoparticle sizes with radii ranging from $1\,\nm$ to $50\,\nm$ to better
understand the physical mechanisms of the proximal and peripheral effects.

Finally, all evaluated target quantities (fluences on the nanoparticle surface,
radial dependencies of frequencies of ionization clusters and energy imparted) are
strictly normalized to the source fluence\footnote{\ Note: Given the two simulation
	steps, confusion may arise regarding terms like `source' and `primary' fluence.
	Unless specified otherwise, these terms will refer to the fluence used in the first
	simulation step and symbols will carry an appropriate index, as delineated in
	Section~\ref{subsection:normalization}.}. This allows establishing a link between
the local effects investigated and macroscopic average dose.

The methodology employed in this work is laid out in Section~\ref{section:methods}.
After detailing the nanodosimetric perspective taken in
Section~\ref{subsection:nanodosimetry}, an account for the two-step simulation
procedure as well as the ionization clustering procedure is given in
Sections~\ref{subsection:simulation} and \ref{subsection:clustering}, respectively.
Section~\ref{subsection:normalization} is dedicated to the normalization to primary
fluence and establishing nomenclature not covered in
Section~\ref{subsection:nanodosimetry}. The results are presented in
Section~\ref{section:results}, followed by a discussion and conclusion in
Section~\ref{section:discussion}.

\section{Methods}
\label{section:methods}

\subsection{Nanodosimetry}
\label{subsection:nanodosimetry}

Conventional dosimetry, focussing on macroscopic averages like absorbed dose (the
mean imparted energy per mass \citep{ICRU_85}), proves reliable on larger scales.
However, target sizes for the initial biophysical radiation action that leads to
cell death have been identified as being of nanometric dimensions
\citep{Goodhead_1994, Goodhead_2006} and damage to cells is largely caused by local
lesion clusters on DNA segments, formed over a few nanometers \citep{Lomax_2013,
	Hill_2020}.

For volumes on this scale, the stochastic nature of radiation interaction dominates
and concepts like local secondary particle equilibrium or the use of an average
energy per ionization to convert from ionization to energy imparted are inapplicable
\citep{Amols_1990}. Therefore, nanodosimetry analyzes the ionization component of
charged particle track structure, the stochastic pattern of ionizing (or exciting)
interactions along a particle trajectory. Classically, nanodosimetry studies the
probabilities of encountering a certain number of ionizations $\nu$, the ionization
cluster size (ICS), within a specific volume at a certain position relative to a
passing track \citep{Grosswendt_2006}.

The central quantity is the probability distribution of ICS, termed ionization
cluster size distribution (ICSD) and denoted as $f_\nu$. Derived metrics such as the
moments of the distribution
\begin{align}
	M_k(Q) \equiv \sum_{\nu\in\mathbb{N}^0} \nu^k f_\nu
\end{align}
and the complementary cumulative frequencies
\begin{align}
	F_k(Q) \equiv \sum_{\nu\ge k} f_\nu,
\end{align}
are extracted from this distribution. For instance, the probability that an
ionization cluster of size \mbox{$\nu\ge 4$} is produced conditional on the occurrence of
an ionization within the sampling volume is $F_4 / F_1$.

Rather than adopting this traditional, probability-based perspective for a single
target volume, this work adopts the conceptionally different perspective of
\citep{Selva_2018}, focussing on total frequency $N_{F_k}$ of clusters of size
\mbox{$\nu\ge k$} per primary fluence $\phi_0$ in a larger, marcroscopic volume (or
their corresponding spatial density):
\begin{align}
	\frac{N_{F_k}}{\phi_0},
\end{align}
with the index $F_k$ alluding to the complementary cumulative frequency identifying the
cluster.

\subsection{Simulation}
\label{subsection:simulation}

The simulation\textquotesingle s aim is to obtain the spatial distribution of
ionization clusters in the proximity of nanoparticles of different sizes ranging
from $1\,\nm$ to $50\,\nm$ irradiated with photons of an energy spectrum
corresponding to the $100$-kVp X-ray spectrum used in a recent multi-center
comparison of Monte Carlo simulations of dosimetric effects of a single gold NP in
water \citep{Li_2020a, Li_2020b}\footnote{\ It should be noted that a further
	(physically implausible) peak between $85.0\,\keV$ and $85.5\,\keV$ has been
	excluded w.r.t. the original spectrum used in \cite{Li_2020a} by averaging over the
	two adjacent bins.}. Some details of the simulation setup used in the preliminary
simulation were also chosen to allow for comparison of the results of this code to
the exercise and follow-up studies \citep{Li_2020a,Li_2020b,Rabus_Li_2021,Rabus_2021}
to be used as a reference.

A realistic track-structure simulation of a gold NP in water is not without
obstacles. Two factors obstruct fast calculations. First, despite gold having a
comparatively high linear attenuation coefficient, a photon traversing a gold
nanoparticle is yet unlikely to interact within it (as an example, the mean number
of photon interactions in a $50\,\nm$-gold-NP ($25\,\nm$ radius) irradiated with
$100$ kVp photons in a beam of $60\,\nm$ can be estimated to $5.4\smalltimes
	10^{-4}$ per photon \citep{Rabus_2021}).

Second, produced electrons may enter or leave the region of interest (here, this is
the NP and the scoring region surrounding it) and accurate simulations need to
properly account for this. The explicit way to ensure this charged particle
equilibrium is to increase the beam size so that the irradiated volume includes all
particles that could possibly be produced and interact within the region of
interest. However, this reduces the incident fluence and many produced primary
particles do not contribute to tallying. In order to achieve acceptable sampling
statistics for quantities differential in shells around the nanoparticle, this
option is considered prohibitively inefficient.

Both of these issues can be addressed by performing the simulation in two separete
steps, referred to as the prior and main simulation.

\subsubsection*{Prior simulation}

In a prior simulation, the fluence of photons and secondary electrons that would be
present on average at any point in a water volume (as a substitute for sub-cellular
matter) where the gold NP would be located is calculated. Fluence, a macroscopic
quantity of course, is scored on finite areas or in finite volumes. To enhance
sampling statistics, the scoring volume was chosen as a cylinder with a radius of
$10\,\cm$ and a height of $100\,\um$, within a larger, cylindrical world volume
(radius of $50\,\cm$ and height of $1\,\mathrm{m}$), both aligned such that the
central axis aligns with the $z$-axis. The wide lateral dimensions of the scoring
and world volumes are to account for photons scattered back into the scoring volume
by Compton or Rayleigh scattering.

For accurate results, track structure calculation was conducted both within the
scoring volume, as well as within $50\,\um$ proximity of the scoring volume to
ensure equilibrium of charged particles with realistic energy spectrum throughout
the entire scoring volume. A distance of $50\,\um$ corresponds to a Continuous
Slowing Down Approximation (CSDA) range of electrons in water with a kinetic energy
of approximately $54.4\,\keV$ \citep{ICRU_37}. As will become evident, roughly
$97.2\,\%$ of the produced secondary electrons have a kinetic energies below that.
Furthermore, the CSDA range is an upper bound for the projected range, which is
typically lower due to the crooked tracks of electrons.

\begin{figure}[t!]
	\fbox{
		\begin{minipage}{\textwidth - 4mm}
			\begin{minipage}{.333\textwidth}
				\subfloat[]{\raggedright{}\label{fig:sim-setup-prior}}\vspace{-0.35cm}
				\begin{center}
					\vspace{2mm}
					\includegraphics[width=\textwidth]{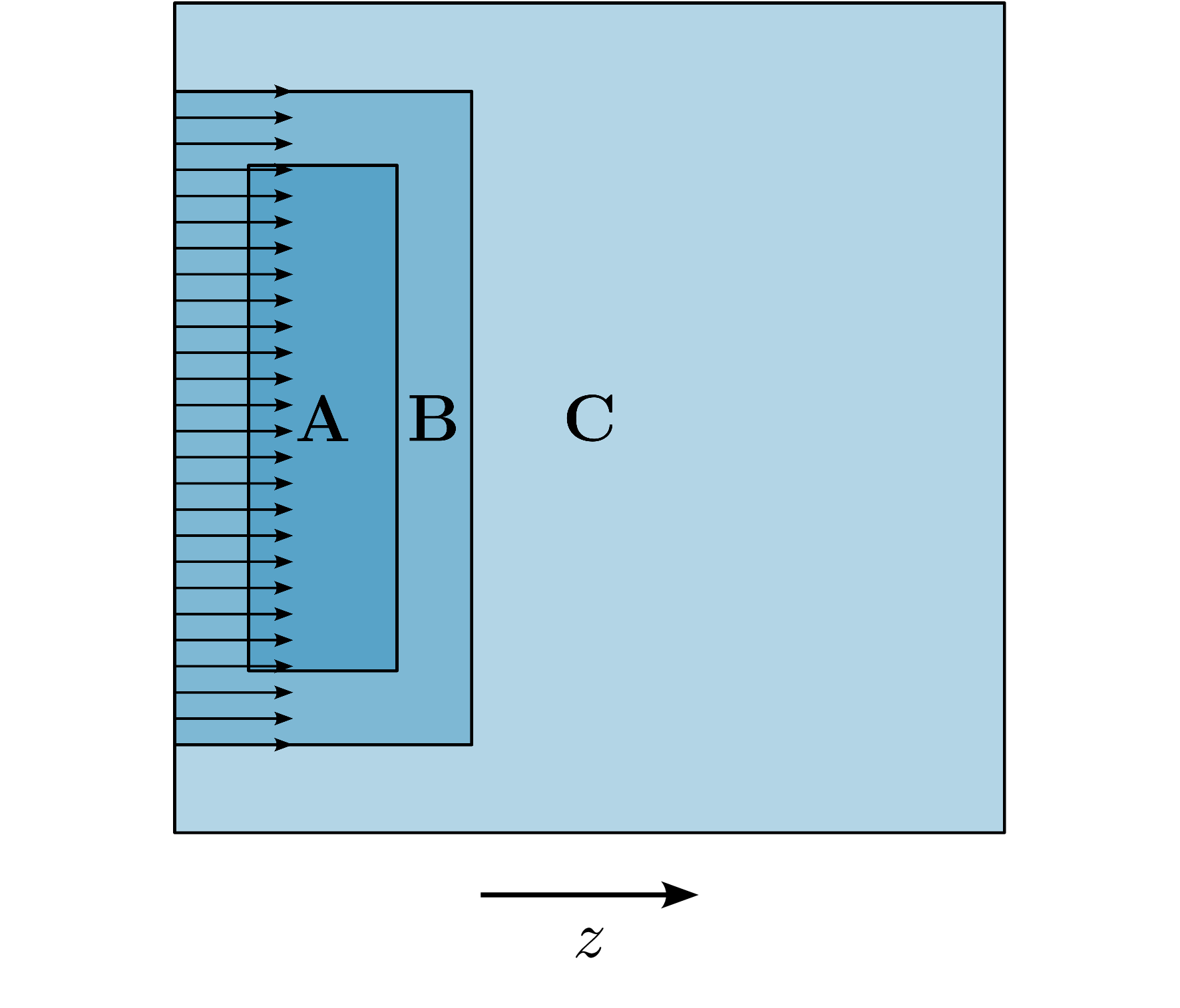}
				\end{center}
			\end{minipage}%
			\begin{minipage}{.333\textwidth}
				\subfloat[]{\raggedright{}\label{fig:sim-setup-main-yx}}\vspace{-0.35cm}
				\begin{center}
					\vspace{2mm}
					\includegraphics[width=\textwidth]{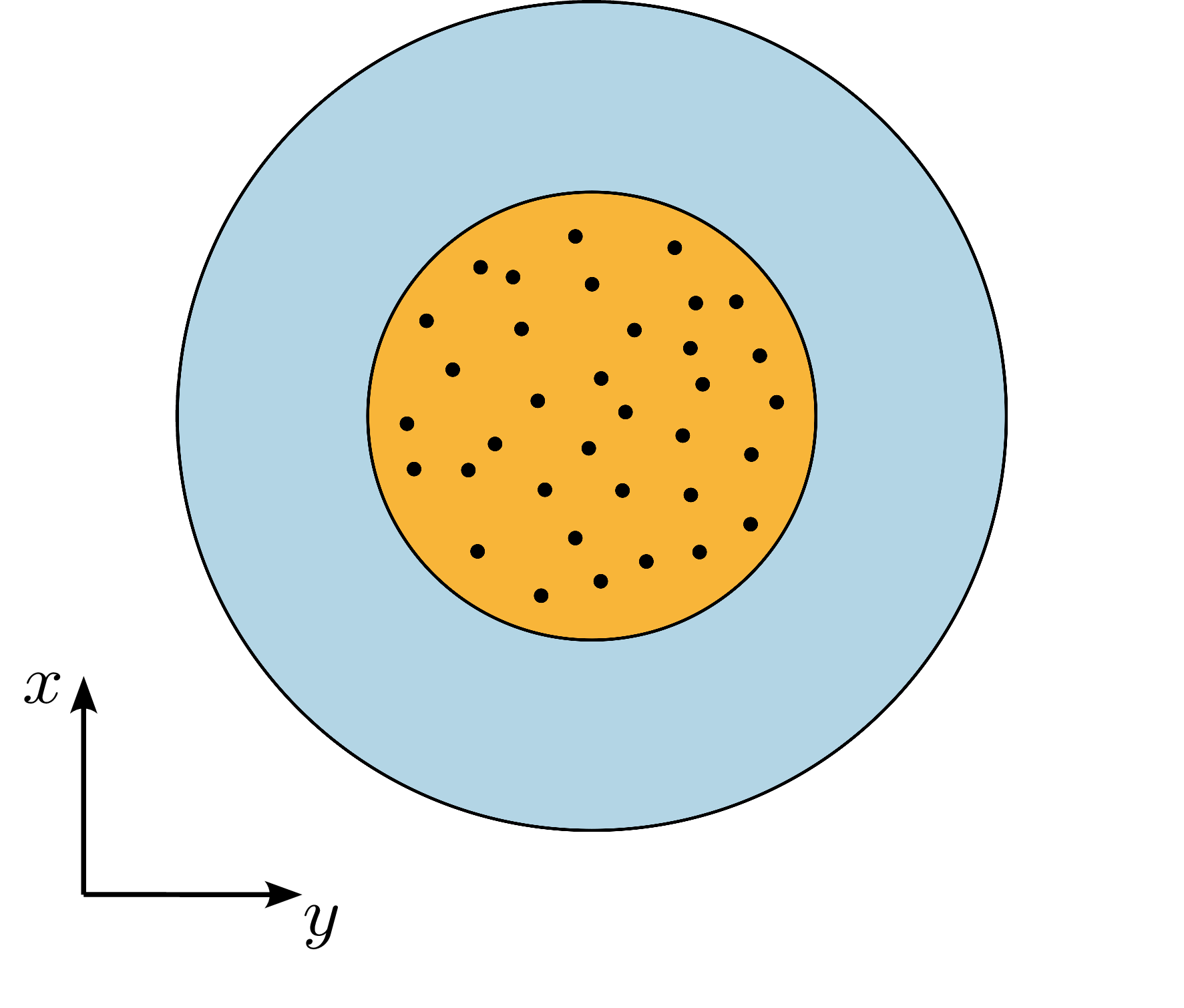}
				\end{center}
			\end{minipage}%
			\begin{minipage}{.333\textwidth}
				\subfloat[]{\raggedright{}\label{fig:sim-setup-main}}\vspace{-0.35cm}
				\begin{center}
					\vspace{2mm}
					\includegraphics[width=\textwidth]{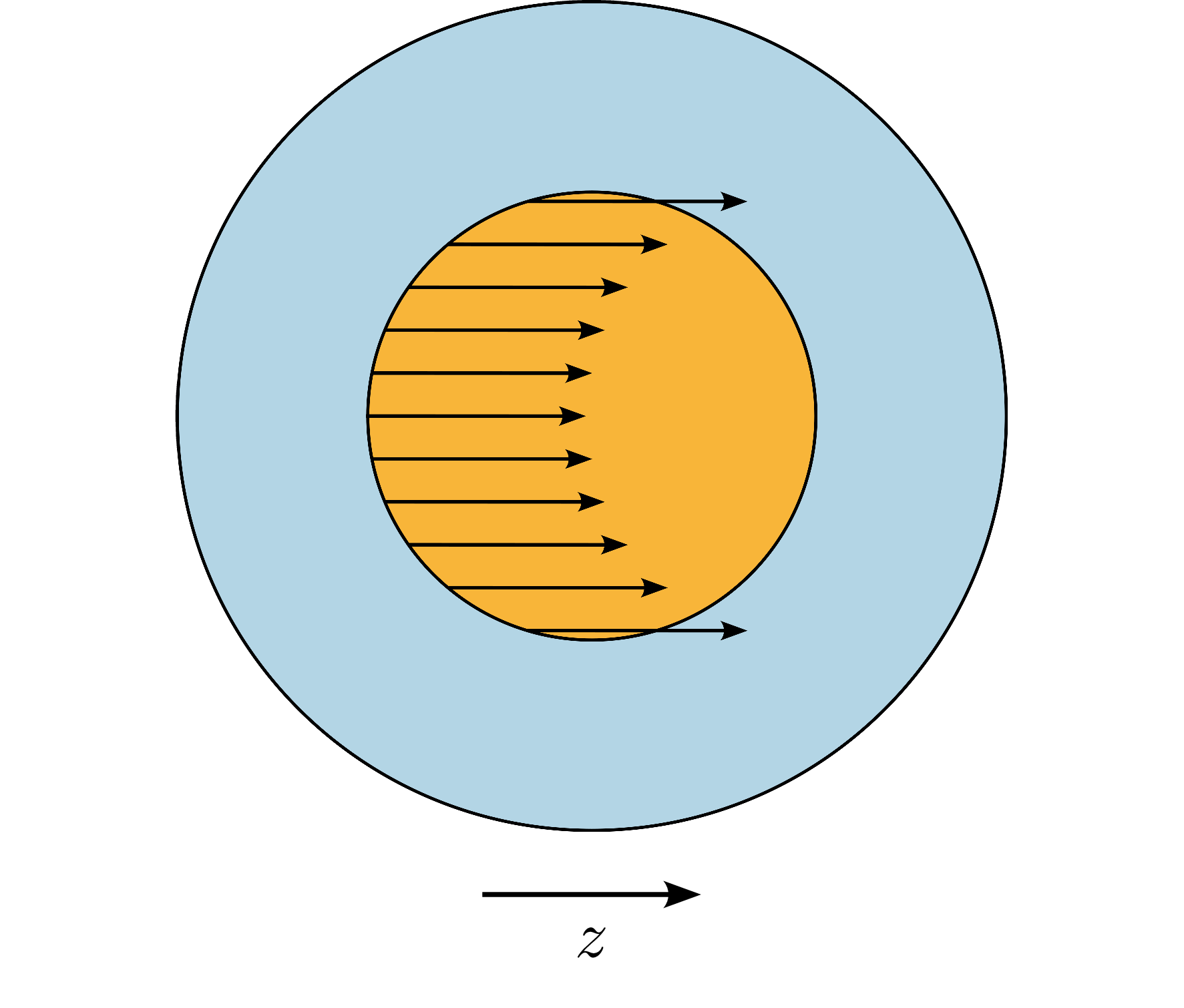}
				\end{center}
			\end{minipage}
			\vspace{-4mm}
			\caption{\textbf{\protect\subref{fig:sim-setup-prior}} Illustration of the
				cylindrical setup of the prior simulation (displayed as the cross-section
				perpendicular to the base) in which a cylindrical scoring volume (region
				`$\mathbf{A}$', radius of \mbox{$10\,\cm$} and height of
				\mbox{$100\,\um$}), situated in the center of an enveloping volume
				(region `$\mathbf{B}$') with a minimum distance between their respective
				surfaces of $50\,\nm$. This enveloping volume is located at the base of
				the world volume (region `$\mathbf{C}$', radius of \mbox{$50\,\cm$} and
				height of \mbox{$1\,\mathrm{m}$}). Track structure calculations are
				performed within the scoring volume as well as the enveloping volume
				(regions `$\mathbf{A}$' and `$\mathbf{B}$').
				Drawing of the irradiation of a gold nanoparticle in water
				from \textbf{\protect\subref{fig:sim-setup-main-yx}} beams-eye-view and
				\textbf{\protect\subref{fig:sim-setup-main}} from the side. Primaries
				are created on one of the NP hemispheres. The nanoparticle is surrounded
				by a spherical scoring volume of water of radius
				$r_\mathrm{roi} = \rnp + 1\,\um$ and a spherical world volume of
				radius $50\,\um$ (not illustrated).
			}
			\label{fig:simulation-setup}
		\end{minipage}
	}
\end{figure}

The inner cylinder is uniformly irradiated along the longitudinal axis (see
Fig.~\subref*{fig:sim-setup-prior}) with photons starting at a distance of
$100\,\um$ from the center of the scoring volume (this corresponds to the interface
of world volume and the volume that envelopes the scoring volume). The beam width
corresponds to the lateral extension of the enveloping volume (region
`$\mathbf{B}$').

Simulations were performed using the Geant4-DNA-library (Version 11.1.1)
\citep{Incerti_2010a, Incerti_2010b, Incerti_2018, Bernal_2015, Sakata_2019}. In
regions `$\mathbf{A}$' and `$\mathbf{B}$', Option 4 models were used for electron
transport below $10\,\keV$ and Option 2 models above. Secondary electrons were not
simulated within region `$\mathbf{C}$'. The step-lengths of the particles within the
scoring volume were tallied to calculate the fluence (as described in
\cite{ICRU_85}), averaged over energy intervals of $100\,\eV$ (for photons) or in
exponentially increasing intervals for electrons as motivated and outlined in
Section~\ref{subsection:fluence_spectra}.

\subsubsection*{Main simulation}

The main simulation (or second simulation) considers a gold nanoparticle in water.
In this simulation, the irradiation source is implicitly implemented by starting
primary particles (photons as well as electrons, separately) on the NP surface with
a distribution of kinetic energies corresponding to the photon and electron fluences
obtained from the prior simulation. Following the approach of \cite{Penelope}, these
energies are sampled using the Walker aliasing algorithm \citep{Walker_1977} for
selection of the energy interval, and the kinetic energy is then uniformly sampled
within this selected interval. This approach is highly efficient and allows for the
generation of primary particle energies from fluence spectra provided as linear or
logarithmic histograms, as done for generated photons and electrons, respectively.

The simulation setup exploits the spherical symmetry of the NP, which allows the
incident particle fluence to be considered isotropic. In this implementation, all
primary particles are generated on one of the NP hemispheres with equal direction of
momentum (say, the $z$-direction) so that the projection of their point of origin
along their initial direction is uniformly distributed within the plane
perpendicular to that direction of momentum (the $x$-$y$-plane, see
Fig.~\subref*{fig:sim-setup-main}). Radiation transport within the gold nanoparticle
was simulated using the models of \cite{Sakata_2019} and the electron transport
in water surrounding the NP is simulated as in the prior simulation.

When performing the clustering of ionizations in a later step, only ionizations
originating from the same primary (of the second simulation) are considered. This
neglects the potential coincident incidence of, for instance, several electrons from
the track of a higher energetic one or of a scattered photon and the related Compton
electron, which are expected to be negligibly low.


In order to determine the net effect of ionization clusters and deposited energies
from a gold NP (see Section~\ref{subsection:normalization}), the main simulation was
also performed with the gold nanoparticle absent and the corresponding volume filled
with water.

\subsubsection*{Scoring}

All simulation results were stored as files in the ROOT format \citep{ROOT}
containing the energy transfer points (namely the event identifier, particle type,
process type, energy deposit and coordinates) as well as step length data
to score fluence (containing the event identifier, particle type, step length and
kinetic energy) in a separate branch of the file. The step length data was used
to score fluence in the prior simulation, whereas the fluence in the main simulation
was determined by scoring only those particles traversing the NP surface and
determining the fluence as the number of such particles per cross-sectional area of
the nanoparticle. The ensuing data analysis (clustering, scoring of imparted energy
etc.) was performed in a pipeline of dedicated \cpp codes. While such data
analysis is often executed at simulation time (i.e. from within the Geant4 user
code), this approach offers the possibility of modifying the data analysis, if so
desired, without having to re-run the simulation. Naturally, this flexibility comes
at the expense of storing the raw data of ionizations and step lengths.

\subsection{Clustering}
\label{subsection:clustering}



\begin{figure}[t!]
	\centering
	\fbox{
		\begin{minipage}{0.667\textwidth}
			\vspace{5 mm}
			\begin{center}
				\includegraphics[width=0.75 \textwidth]{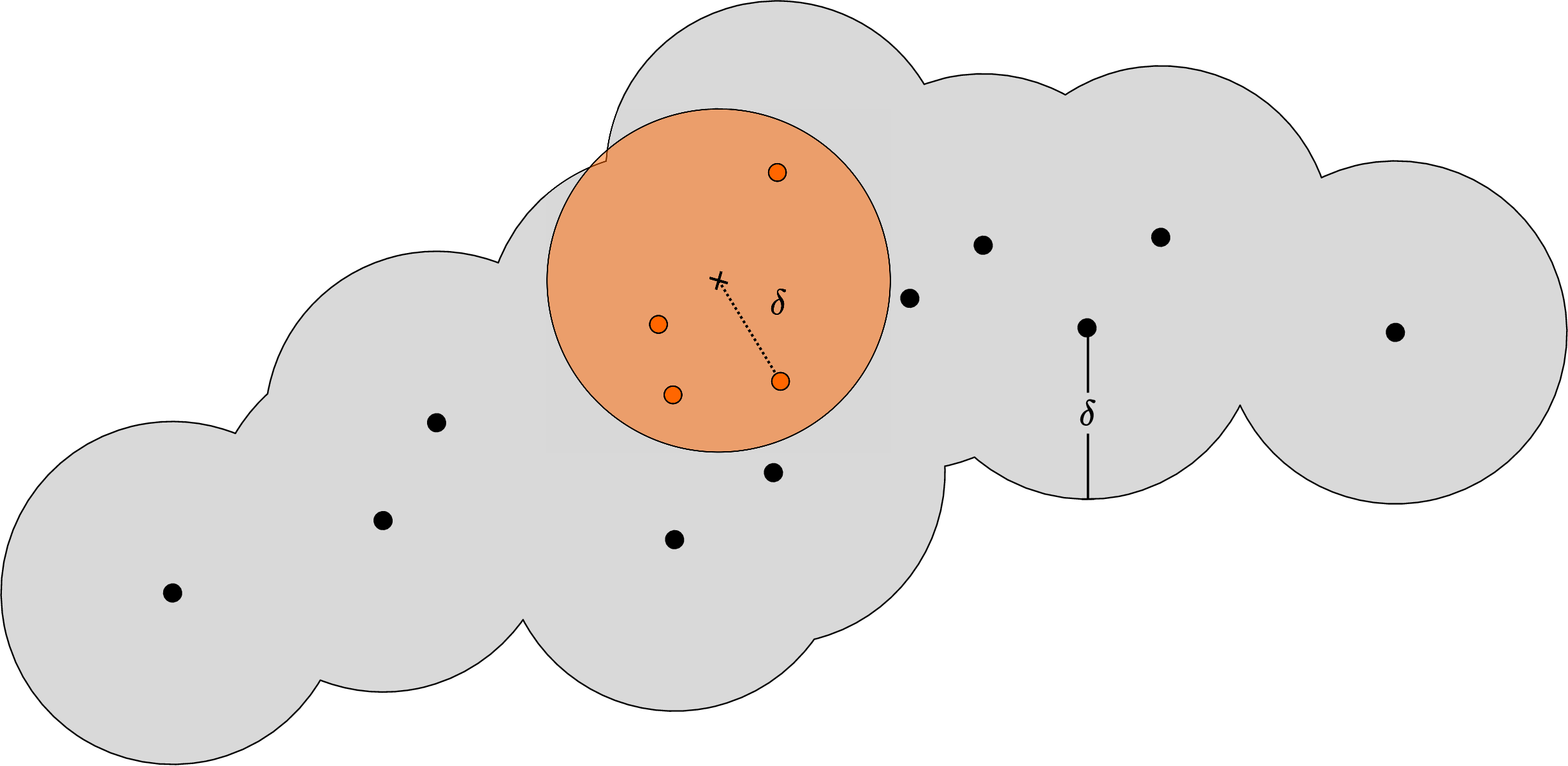}
			\end{center}
			\caption{Illustration of the Associated Volume Clustering approach. The points
				symbolize the ionization sites the algorithm iterates through.
				The union of spheres with radius $\delta$ is referred to as the
				associated volume of the track (gray area).
				For each site, a random point is sampled within a sphere of radius
				$\delta$. All points within a distance $\delta$ from this point
				(orange area) are then considered to be on the same cluster, which is
				then scored with a weight of $1/\nu$, where $\nu$ is the cluster
				size (see text for details).
			}
			\label{fig:avc}
		\end{minipage}
	}
\end{figure}

In this work, ionization clusters are scored in spherical targets of
\mbox{$1.5\,\nm$} radius, which are equal in volume to the often used cylinders with
a height of \mbox{$3.4\,\nm$} and radius of \mbox{$1.15\,\nm$} representing a DNA segment of
ten base pairs \citep{Grosswendt_2006}. The sampling volumes are not uniformly
distributed in the entire region of interest (ROI) as, for instance, in
\cite{Alexander_2015} and \cite{Ramos-Mendez_2018} as this is inefficient, especially for
loosely ionizing particles such as photons. Instead, they are sampled within a
volume enveloping the track, the track’s associated volume (AV) \citep[Chapter 2, by
	A. M. Kellerer]{Kase_Bjarngard_Attix_1985}. The associated volume of a track (gray
area in Fig.~\ref{fig:avc}) is the union of all spheres of radius $\delta$ centered
at the loci of all $n_t$ energy transfer points with ionizations. The AV of the
track, $V_t$, and the AVs of the individual ionizations, $V_i$, can be segmented in
disjoint sub-volumes, $V_{t,\nu}$ and $V_{i,\nu}$, respectively, representing the
intersections of $\nu$ different spheres such that

\begin{align}
	V_t = \sum_\nu V_{t,\nu}\,,
	\quad V_i = \sum_\nu V_{i,\nu}
	\quad\mathrm{and}\quad V_{t,\nu} = \frac1\nu\sum_{i=1}^{n_t} V_{i,\nu}\,.
	\label{eq:volumes}
\end{align}

The normalization to $\nu$ in the last identity in Eq.~\ref{eq:volumes} accounts for
the fact that each such intersection is counted $\nu$ times in the sum. Averaged
over tracks, the $V_{t,\nu}$ are proportional to the probability of formation of an
ionization cluster of size $\nu$. The AV clustering algorithm cycles through the
ionization sites of a track. For each site, a point is sampled within a sphere of
radius $\delta$. A scoring volume of the same size (orange area in
Fig.~\ref{fig:avc}) is placed at this point and all ionization sites within this
volume are counted to establish the ionization cluster size $\nu$, which is then
counted with a weight of $1/\nu$ to account for the (potential) $\nu$-fold counting.

\subsection{Notation and Normalization}
\label{subsection:normalization}

\subsubsection*{Notation}

Important quantities are the fluences of the appearing radiation fields, such as the
(integral) primary fluence $\phi_0$, defined as the number of particles $\dd N_0$
incident on a sphere of cross-sectional area $\dd A$ \citep{ICRU_85}
\begin{align}
	\phi_0 = \frac{\dd N_0}{\dd A}\label{eq:fluence}
\end{align}
or its corresponding distribution with respect to energy $\dd
	\phi_0 / \dd E$ (referred to as spectral fluence).

The two-step simulation approach readily allows individuating the different
components of the radiation field to clustering as well as energy imparted, namely
the four contributions of a primary photon causing either a photon or an electron to
enter the NP (simulation I) and upon interaction a photon or an electron to leave
the NP (simulation II):
\begin{align*}
	\gamma		  \overset{\mathrm{I}}\longrightarrow \{\gamma, e^-\}
	 & \overset{\mathrm{II}}\longrightarrow \{\gamma, e^-\}
\end{align*}
Correspondingly, the fluences are denoted as
\begin{align*}
	\phi_0^{(\gamma)} \overset{\mathrm{I}}\longrightarrow \phi_1^s
	 & \overset{\mathrm{II}}\longrightarrow
	\phi_2^{s\to s^\prime}
\end{align*}
with $s, s^\prime \in \{\gamma, e^-\}$. Namely, $\phi_1^s$ denotes the fluence of
particles of type $s$ (entering the nanoparticle) and $\phi_2^{s\to s^\prime}$
denotes the fluence of particles of type $s^\prime$ (leaving the nanoparticle),
produced by a particle of type $s$ (entering the nanoparticle).

In order to link local effects around the nanoparticle to macroscopic quantities,
such contributions need to be normalized meaningfully, for instance, to the primary
fluence $\phi_0$. For the spectral fluence penetrating the nanoparticle $\dd\phi_1^s
	/ \dd E$, this is straight-forwardly
\begin{align*}
	\frac{1}{\phi_0} \frac{\dd \phi_1^s}{\dd E}
\end{align*}

For the second simulation, this is non-trivial as different numbers of histories are
simulated for photons and electrons, which do not correspond to the primary fluence.
Here, the notation will distinguish between the ``actual'' fluences $\phi_i$, which
correspond to the primary radiation field in the first simulation and the fluences
used in the second simulation, marked with a tilde $\tilde\phi_i$ (in the case of
primary fluence, these two quantities simply coincide, i.e. $\phi_0 =
	\tilde\phi_0$). The fluence $\phi_2^{s\to s^\prime}$ depends on the fluence
$\phi_1^s$ via $p^{s \to s^\prime}(E^\prime|E)$, the expectation of the number of
particles of type $s^\prime$ and energy $E^\prime$ that are produced when a particle
of type $s$ with energy $E$ enters the nanoparticle, so that:
\begin{align}
	\begin{split}
		\frac{1}{\phi_0} \frac{\mathrm{d} \phi_2^{s \to
				s^\prime}}{\mathrm{d} E^\prime}
		 & = \frac{1}{\phi_0} \int\mathrm{d}E\ p^{s \to
				s^\prime}(E^\prime|E)\,\,
		\frac{\mathrm{d} \phi_1^{s}}{\mathrm{d} E}
		\\
		 & = \frac{1}{\phi_0} \frac{\phi_1^s}{\tilde\phi_1^s}\int\mathrm{d}E
		\ p^{s \to s^\prime}(E^\prime|E)\,\,
		\frac{\mathrm{d} \tilde\phi_1^{s}}{\mathrm{d} E}
		\\
		 & = \frac{1}{\phi_0} \frac{\phi_1^s}{\tilde \phi_1^s}
		\frac{\mathrm{d}
			\tilde\phi_2^{s \to s^\prime}}{\mathrm{d} E^\prime}
	\end{split}
\end{align}

The same rationale can be applied for a generic quantity of interest $Q$ (here as a
linear density, as a distribution of $x$, for example the radial distance $r$ from
the nanoparticle center), resulting in the same normalization factor, so that:

\begin{align}
	\frac{1}{\phi_0} \frac{\dd Q}{\dd x} = \frac{1}{\phi_0}
	\frac{\phi_1^s}{\tilde\phi_1^s} \frac{\dd \tilde Q}{\dd x} \label{eq:QOI}
\end{align}

To account for the expected background contribution to a quantity of interest, the
net contribution from the gold nanoparticles must be calculated, namely the surplus
effect that the presence of the nanoparticle adds compared to when it is absent. For
this purpose the main simulation was performed with a spherical water volume, as a
stand-in for the nanoparticle, as described in Section~\ref{subsection:simulation}.
The difference between the gold NP contribution $Q_\mathrm{g}^{s\to s^\prime}(\rnp)$
and the corresponding, spherical water volume $Q_\mathrm{w}^{s\to s^\prime}(\rnp)$
with $s, s^\prime\in\{\gamma,e^-\}$ for a contribution is called the net
contribution
\begin{align}
	\frac{\dd Q_\mathrm{net}^{s\to s^\prime}}{\dd x}(\rnp) \equiv
	\frac{\dd Q_\mathrm{g}^{s\to s^\prime}}{\dd x}(\rnp)
	- \frac{\dd Q_\mathrm{w}^{s\to s^\prime}}{\dd x}(\rnp),
	\label{eq:net-contribution}
\end{align}
to which the background contribution from the prior simulation is then added:
\begin{align}
	\frac{1}{\phi_0}\frac{\dd Q}{\dd x}(\rnp)
	= \sum_{s, s^\prime\in\{\gamma,e^-\}}
	\!\frac{1}{\phi_0}\frac{\dd Q_\mathrm{net}^{s\to s^\prime}}{\dd x}(\rnp)
	+ \frac{1}{\phi_0}\frac{\dd Q^\mathrm{b}}{\dd x}
	\label{eq:background-correction}
\end{align}

where $r$ is the gold NP radius and $\dd Q^\mathrm{b}$ is the background
contribution within the interval $[x, x + \dd x)$. In case that $x=r$, the radial
distance from the nanoparticle center, $\dd Q^\mathrm{b}(r) = Q^\mathrm{b} /
	V_\mathrm{b} \cdot \dd V(r)$ (with $V_\mathrm{b}$ the scoring volume of the prior
simulation) is the contribution from the background simulation, in a volume of the
spherical shell comprised within the interval $[r, r + \dd r)$.

\subsubsection*{Conditional normalization}

The quantities determined by the identity on the right-hand side of Eq.
\ref{eq:background-correction} are averaged over all primary particles of the
respective second simulation. For the case of incident photons, the probability of
an interaction in the gold NP is relatively low. Therefore, it is interesting to
look at the distribution of ionization clusters around a gold NP in which an
ionizing interaction of a photon occurs. In most cases, the incident photon does not
interact in the gold NP, no electrons are emitted from the gold NP and the expected
cluster density around the gold NP is equal to the background contribution. If the
photon interacts, the cluster density is enhanced with respect to the average
cluster density obtained from the simulation. The enhancement is by a factor of
$1/\bar n_0^\gamma(\rnp)$, where $\bar n_0^\gamma(\rnp) \equiv \bar n^\gamma(\rnp,
	\phi_0)$ is the expected number of photon interactions taking place in the gold NP
with radis $\rnp$ for a primary photon fluence $\phi_0$.

The probability of an incident photon undergoing an ionizing interaction in the gold
NP is approximately (in lowest order approximation) given by
\begin{align}
	\begin{split}
		\big<p_i^\gamma \big>_{\phi_1^\gamma} & \ =\
		\bar\ell\ \big<\mu_{i,\mathrm{g}}\big>_{\phi_1^\gamma}
		\ =\ \frac{\bar\ell}{\phi_1^\gamma} \int\!\dd E\, \frac{\dd
			\phi_1^\gamma}{\dd E}\,\mu_{i,\mathrm{g}}(E)
		\label{eq:p_g}
	\end{split}
\end{align}

Here, $\bar\ell = 4 \rnp/3$ is the mean chord length (or path length) of a photon
traversing the (spherical) nanoparticle, $\mu_{i,\mathrm{g}}$ denotes the linear
attenuation coefficient of gold for ionizing interactions, and $\dd \phi_1^\gamma /
	\dd E$ is the energy spectrum of the photons impinging on the gold NP. The photon
interaction probabilities used in this work are obtained using this method. The
values used of the linear attenuation coefficients are from the generalized
spline-interpolation of the data from \cite{Berger_2010}, presented in
\cite{Rabus_2019}. The probability of a photon interaction can also be estimated
from the fluence spectra obtained in the second simulation as follows:
\begin{align}
	\big<p_i^\gamma\big>_{\phi_1^\gamma} & \ =\ 1 -
	\frac{\phi_2^{\gamma\to\gamma}(E_\mathrm{min})}{\phi_1^{\gamma}(E_\mathrm{min})}
	\qquad\mathrm{with}\qquad
	\phi_i(E_\mathrm{min}) \equiv \int_{E_\mathrm{min}}^\infty\!\mathrm{d}E\,\frac{\dd\phi_i}{\dd E}
	\label{eq:p_g_sim}
\end{align}

Here $E_\mathrm{min}$ was chosen as $15\,\keV$, which is an effective lower bound of
the relevant section of the photon spectrum and excludes low-energy fluerescence
photons produced within the gold NP . The two approaches to determine interaction
probability (Eq.~\ref{eq:p_g} and Eq.~\ref{eq:p_g_sim}) yield slightly different
results, which reflect the uncertainty of the cross-sections \cite{Andreo_2012}. A
comparison of the probabilities obtained from both methods for the nanoparticle
radii used in this work can be found in Supplementary Table~\ref{tab:p_vs_p}.

The expected number of ionizing interactions in gold per primary fluence $\phi_0$ is
then given by
\begin{align}
	\frac{\bar n_0^\gamma(\rnp)}{\phi_0}\ =\
	\rnp^2\pi\ \frac{\phi_1^\gamma}{\phi_0}\ \big<p_i^\gamma \big>_{\phi_1^\gamma}
	\ =\ \frac{4\pi}3\rnp^3\frac{\phi_1^\gamma}{\phi_0}
	\ \big<\mu_{i,\mathrm{g}}\big>_{\phi_1^\gamma} \ \propto\ \rnp^3    \label{eq:n-bar_per_phi}
\end{align}
and is proportional to the volume of the gold NP. When a quantity of interest as
determined by Eq.~\ref{eq:QOI} is normalized to the ratio given in
Eq.~\ref{eq:n-bar_per_phi}, the result is the expected value of this quantity of
interest conditional on the occurrence of a single ionizing interaction of a photon
occurring in the gold NP and does not depend on the primary fluence. If the gold NP
undergoes $k$ ionizations by photons, the expected values of the conditional
quantity are $k$ times higher, again independent of primary fluence.

However, the expected number of ionizing photon interactions in a gold NP depends on
the primary fluence. Therefore, the values of a quantity of interest conditional on
at least one ionizing interaction occurring in the gold NP depends on the primary
fluence, with different primary fluences corresponding to different ratios of the
contribution of the gold NP undergoing an interaction to the background
contribution. A natural choice seems to be values of primary fluence $\phi_0^D$ that
correspond to given values of absorbed dose $D$, given by
\begin{align}
	\phi_0^D\ \equiv\ D\frac{\phi_0}{D_0}
\end{align}
with $D_0$ being the dose corresponding the the primary fluence $\phi_0$. The
expected number of interactions per fluence in Eq.~\ref{eq:n-bar_per_phi} is
independent of fluence, so that
\begin{align}
	\frac{\bar n_0^\gamma}{\phi_0} = \frac{\bar
		n_D^\gamma}{\phi_0^D}\quad\Longrightarrow\quad
	\bar n_D^\gamma \ \equiv\ \bar n^\gamma(D)\ =\ \phi_0^D \frac{\bar
		n_0^\gamma}{\phi_0}. \label{eq:n_D}
\end{align}
In Eq.~\ref{eq:n_D} and in the following, the argument $\rnp$ is dropped to abridge
notation.

Even though, the probability of a single photon undergoing an ionization event
within a gold nanoparticle is relatively low, the number of such events occurring in
a single nanoparticle aggregates for fluences corresponding to higher doses. Since
interactions of different photons can be considered statistically independent, the
probability of the number of interactions occurring for an expected number of
interactions $n_D^\gamma$ is Poisson-distributed and the probability of $k$ events
occurring is thus given by
\begin{align*}
	P(k|n_D^\gamma) = \frac{(\bar n_D^\gamma)^k}{k!}\mathrm{e}^{-\bar n^\gamma}
\end{align*}
and the probability of one or more ionizing events sums up to
\begin{align}
	P(k \ge 1|n_D^\gamma) = \mathrm{e}^{-\bar n_D^\gamma} \sum_{k\in\mathbb{N}}\frac{(\bar n_D^\gamma)^k}{k!}
	= 1 - \mathrm{e}^{-\bar n_D^\gamma}.    \label{eq:p_ge1}
\end{align}

A quantity of interest $Q$ conditional on one or more interactions occurring in a
gold NP is, therefore, given by Eq.~\ref{eq:dQdx_C}. The factor in front of the
curly brackets is the expectation of the number of interactions conditional on the
occurrence of at least one.
\begin{align}
	\frac{\dd Q_\mathrm{C}^\gamma}{\dd x}(D)
	\ =\ \frac{\bar n_D^\gamma}{1 - \mathrm{e}^{-\bar n_D^\gamma}}
	\left\{\frac1{\bar n_0^\gamma}\frac{\dd Q_\mathrm{net}^{
			\gamma\to\gamma}}{\dd x} + \frac1{\bar n_0^\gamma}\frac{\dd
		Q_\mathrm{net}^{\gamma\to e}}{\dd x}\right\}
	+ \frac{\phi_0^D}{\phi_0}\frac{\dd Q^\mathrm{b}}{\dd x}
	, 	\label{eq:dQdx_C}
\end{align}

\subsubsection*{Relevant quantities}

Quantities of interest are the number of clusters of size $\nu \ge k$,
$N_{F_k}$, as a function of radial distance $r$ and the number of such clusters per
mass, a quantity corresponding to the cluster dose $g^{F_k}$, introduced in
\cite{Faddegon_2023}, as well as the energy $\varepsilon$ imparted by ionizations
and the corresponding dose contribution $D$ from ionizations. A summary of the
relevant quantities can be found in Supplementary Table~\ref{tab:quantities}.


\section{Results}
\label{section:results}

\subsection{Fluence spectra}
\label{subsection:fluence_spectra}

\begin{figure}[t!]
	\fbox{
		\begin{minipage}{\textwidth - 4mm}
			\begin{minipage}{.5\textwidth}
				\subfloat[]{\raggedright{}\label{fig:fluence_prior_g}}\vspace{-0.35cm}
				\begin{center}
					\includegraphics[width=\textwidth]{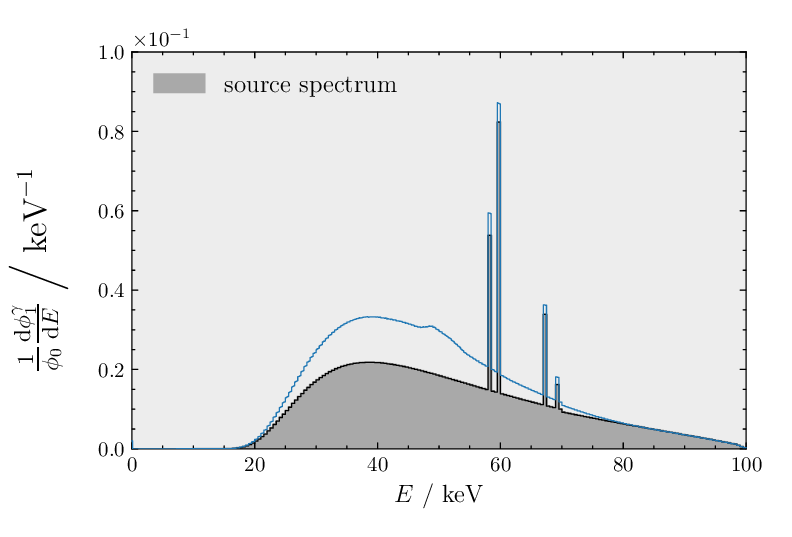}
				\end{center}
			\end{minipage}%
			\begin{minipage}{.5\textwidth}
				\subfloat[]{\raggedright{}\label{fig:fluence_prior_e}}\vspace{-0.35cm}
				\begin{center}
					\includegraphics[width=\textwidth]{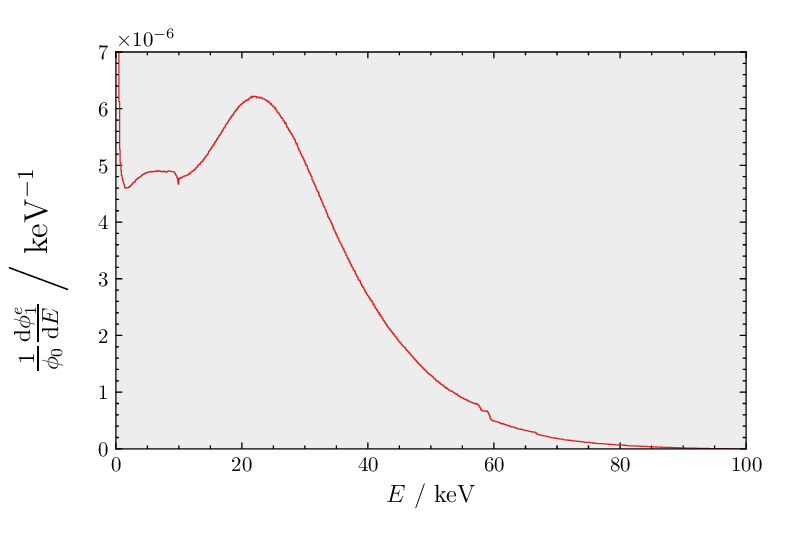}
				\end{center}
			\end{minipage}
			\caption{\textbf{\protect\subref{fig:fluence_prior_g}} Spectral fluences of the
				photons emitted from the $100$ kVp X-ray source (gray shaded) and of the
				resulting photon field in the scoring volume (blue line), normalized to the
				integral photon fluence from the source (primary fluence).
				The bin width is $100\,\eV$.
				The latter photon spectrum noticeably differs from the source spectrum due
				to photons scattering back into the scoring volume via incoherent
				scattering (Compton effect) as well as coherent scattering (Rayleigh
				scattering). \\
				\textbf{\protect\subref{fig:fluence_prior_e}} Resulting electron
				fluence in the scoring volume normalized to the
				primary photon fluence. Note that fluence values for energies  $E<500\,\eV$ have
				been cut for legibility. The fluence reaches values up to
				\mbox{$8.3\smalltimes10^{-4} \,\keV^{-1}$} at these energies; see
				Fig.~\protect\subref*{fig:fluence_ee_log} for a more detailed view of
				the electron fluence at lower energies.
			}
			\label{fig:prior_fluences}
		\end{minipage}
	}
\end{figure}

The fluence data obtained from the prior simulation is depicted in
Fig.~\ref{fig:prior_fluences}. The photon fluence in the region of interest is
overall higher than that of the primary spectrum ($\phi_1^\gamma \approx 1.38
	\,\phi_0$). This is a consequence of Compton as well as Rayleigh scattering of
photons in the large simulation volume back into the region of interest (ROI). The
secondary-electron fluence is significantly lower than the photon fluence ($\phi_1^e
	\approx 3.18 \smalltimes 10^{-4} \,\phi_1^\gamma$) and characterized by a continuum
of photo-electrons as well as lower energetic Compton electrons at energies
$\lesssim 10\,\keV$.

\begin{figure}[t!]
	\centering
	\fbox{
		\begin{minipage}{0.667\textwidth}
			\begin{center}
				\includegraphics[width=0.75
					\textwidth]{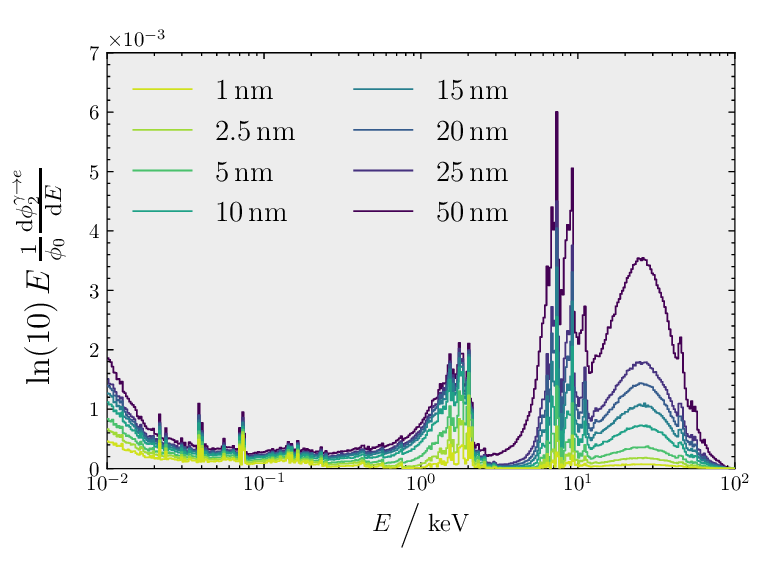}
			\end{center}
			\vspace{-5mm}
			\caption{Spectral energy fluence of emitted electrons (red line) produced by
				photons, normalized to the primary fluence for different nanoparticle
				radii (colored lines).
				Note the logarithmic $x$-axis and linear $y$-axis. In this
				representation,
				the area under the curve in an interval is the number of electrons per
				primary particle in that interval.
			}
			\label{fig:surface_fluence_ge}
		\end{minipage}
	}
\end{figure}

\begin{figure}[t!]
	\fbox{
		\begin{minipage}{\textwidth - 4mm}
			\begin{minipage}{.5\textwidth}
				\subfloat[]{\raggedright{}\label{fig:fluence_ee}}\vspace{-0.35cm}
				\begin{center}
					\includegraphics[width=\textwidth]{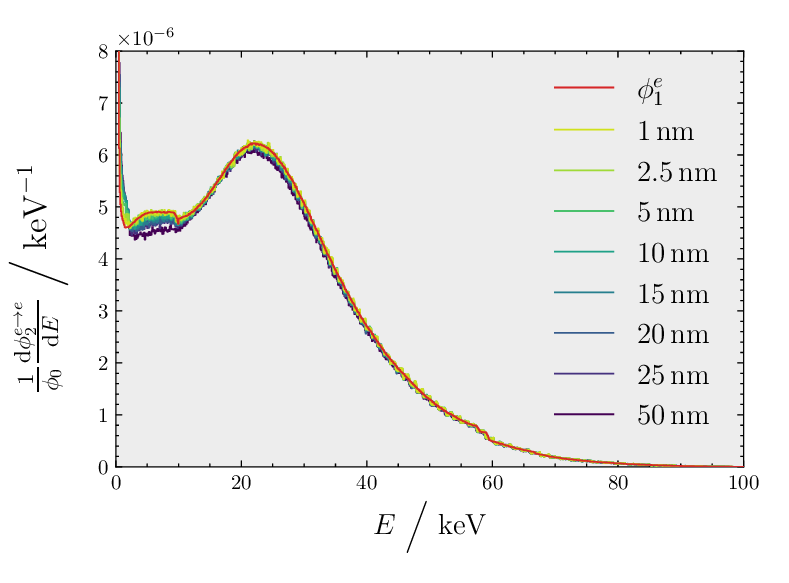}
				\end{center}
			\end{minipage}%
			\begin{minipage}{.5\textwidth}
				\subfloat[]{\raggedright{}\label{fig:fluence_ee_log}}\vspace{-0.35cm}
				\begin{center}
					\includegraphics[width=\textwidth]{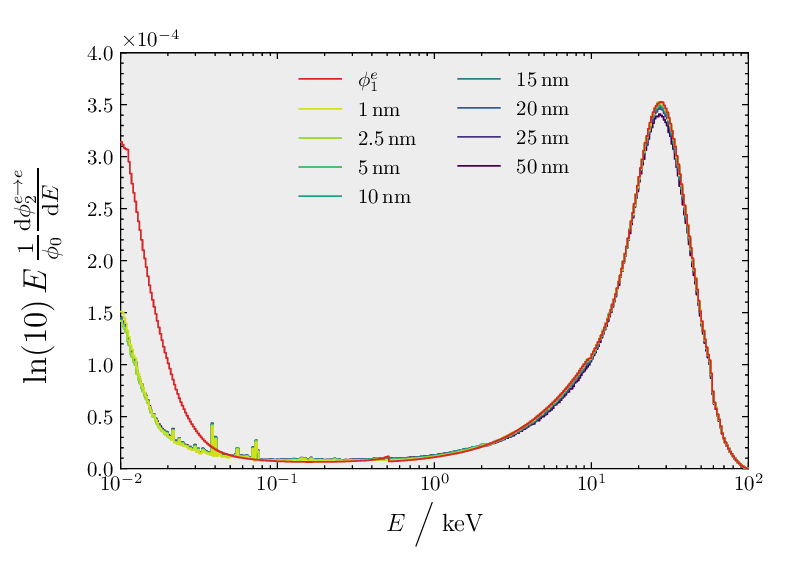}
				\end{center}
			\end{minipage}
			\caption{
				Comparison of the distribution of the fluence of incident electrons
				with respect to energy and the corresponding fluence spectra of
				electrons emitted from the gold nanoparticle. All data is normalized to the primary
				photon fluence. \textbf{\protect\subref{fig:fluence_ee}} Spectra obtained
				by linear binning of the energy of emitted electrons with the same
				\mbox{$100\,\eV$} bin size as for the incident electrons. As in
				Fig.~\protect\subref*{fig:fluence_prior_e}, the data at energies below
				\mbox{$500\,\eV$} are outside the range shown on the vertical axis.
				\textbf{\protect\subref{fig:fluence_ee_log}} Spectra of emitted electrons
				using energy binning with equidistant logarithmic intervals. In this case,
				the $y$-axis has been multiplied by a factor $\mathrm{ln}(10)E$, so that the
				area under the curve is representative of the number of electrons in the
				respective energy interval.
			}
			\label{fig:surface_fluence_ee}
		\end{minipage}
	}
\end{figure}

By virtue of the two-step procedure in simulation, the single contributions to the
fluence leaving the nanoparticle are readily available and allow gaining a clearer
perspective on the different physical processes governing the formation of
ionization clusters in the nanoparticle's periphery. Most relevant to the formation
of clusters are the fluences of emitted electrons produced by incident photons or
electrons. Fig.~\ref{fig:surface_fluence_ge} and Fig.~\subref*{fig:fluence_ee_log}
report the energy distribution of electrons on the surface of nanoparticles with
different radii, distinguishing between those of electrons produced by photons
$\phi_2^{\gamma\to e}$ and by electrons $\phi_2^{e\to e}$ entering the nanoparticle.
The binning is logarithmic with a minimum energy of $E_\mathrm{min} = 10\,\eV$, and
$100$ bins per decade, so that the positions of the lower bin edges are given by
\begin{align*}
	\Big\{E_\mathrm{min} \cdot 10^{k/100}\ \big|\ k\in\mathbb{N}^0\Big\}.
\end{align*}

The data are plotted in `microdosimetry style', whereby the linear vertical axis
shows the fluence multiplied with the $\mathrm{ln}(10)E$. While technically the
quantity shown on the $y$-axis is proportional to the energy fluence, this way of
presentation assures that the area under the curve in any energy interval is
representative of the contribution of electrons in this interval to the total number
of emitted electrons.

The photon-induced electron spectrum in Fig.~\ref{fig:surface_fluence_ge} exhibits
an overall dependence on nanoparticle radius $\rnp$. In fact, the number of
electrons leaving the nanoparticle $N_2^{\gamma\to e}$ is proportional to the mean
number of photon-interactions $\bar n_0^\gamma(\rnp)$ within the nanoparticle,
introduced in Eq. \ref{eq:n-bar_per_phi}, so that
\begin{align*}
	\phi_2^{\gamma\to e}\ =\ \frac{N_2^{\gamma\to e}}{A_\mathrm{np}}\ \propto\ \rnp.
\end{align*}

Qualitatively, the spectra in Fig.~\ref{fig:surface_fluence_ge} contain a continuum
of photo-electrons, mainly at energies $E\gtrsim 10\,\keV$, and a number of
partially overlapping peaks that correspond to Auger transitions. For the binning
used, it may not be possible to identify all peaks individually at larger energies
as two separate peaks might coalesce. Additionally, the Auger electrons are subject
to some straggling on their way to the nanoparticle surface, which smears the peaks
to some extent.


The prominent peaks at energies between $7.3$-$7.4\,\keV$, $9.2$-$9.3\,\keV$ as well
as between $11.1$-$11.2\,\keV$ are caused by electrons from L shell Auger
transitions\footnote{\ The gold models of Geant4 \citep{Sakata_2019} used here use
	transition energies from the EADL database \mbox{\citep{EADL}}.}. Further peaks at
energies from $1.5\,\keV$ up to $2.1\,\keV$ originate from M shell Auger electrons
and the peaks at lower energies can likely be associated with N or
O shell transitions or Coster-Kronig transitions.


Qualitatively, the electron-induced electron fluence spectra $\phi_2^{e\to e}$ shown
in Fig.~\subref*{fig:fluence_ee} appear almost the same as the spectrum of electrons
entering the nanoparticle $\phi_1^e$. A minor decrease of the spectral fluence can
be seen at energies above $2\,\keV$, which is most pronounced in the energy range
from $2\,\keV$ to $10\,\keV$ and can be attributed to impact ionization within the
NP (for larger radii). However, the integral fluences are almost independent of
radius.

Since the probability of incident electrons undergoing inelastic interactions in the
gold NP decreases with decreasing gold NP size, it is expected that the energy
spectra leaving the gold NP asymptotically approach the incident electron spectrum
with decreasing radii. While such behavior can be observed at higher energies in
both Fig.~\subref*{fig:fluence_ee} and Fig.~\subref*{fig:fluence_ee_log}, the
fluence spectra of emitted electrons visibly deviate from the spectral fluence of
incident electrons at energies below 100 eV.

This behavior is attributed to the high probability of low energy electrons to
undergo an inelastic interaction in gold (see Supplementary
Fig.~\ref{fig:p_penelope}). While impact ionization by higher energy electrons
predominantly leads to a production of electrons at low energies, the high
probability of inelastic interactions at low energies effectively leads to an
absorption of these electrons (or, more precisely, deceleration to energies below
the ionization threshold of water).

In comparison, the \mbox{$(\gamma\!\to\! e)$\,-\,component} dominates the electron
fluence on the gold NP surface especially for growing nanoparticle sizes; the value
of $\phi_2^{\gamma\to e} / \phi_2^{e\to e}$ grows from $1.43$ ($\rnp = 1\,\nm$) to
$16.94$ ($\rnp = 50\,\nm$). For integral fluences at energies corresponding to
M shell Auger electrons ($500\,\eV \le E < 5.5\,\keV$), this fraction ranges from
$2.90$ ($\rnp = 1\,\nm$) to $34.98$ ($\rnp = 50\,\nm$). See Supplementary
Table~\ref{tab:phi_ge_ee} for a more detailed overview.

\subsection{Ionization clustering}
\label{subsection:ionization_clustering}

\begin{figure}[t!]
	\fbox{
		\begin{minipage}{\textwidth - 4mm}
			\begin{minipage}{.5\textwidth}
				\subfloat[]{\raggedright{}\label{fig:ID_all}}\vspace{-0.35cm}
				\vspace{-0.2mm}
				\begin{center}
					\includegraphics[width=\textwidth]{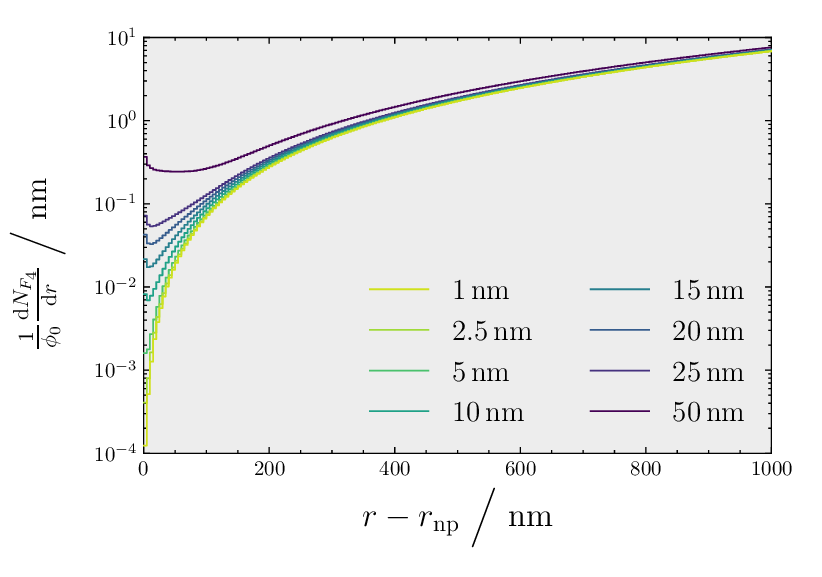}
				\end{center}
			\end{minipage}%
			\begin{minipage}{.5\textwidth}
				\subfloat[]{\raggedright{}\label{fig:g_all}}\vspace{-0.35cm}
				\begin{center}
					\includegraphics[width=0.98\textwidth]{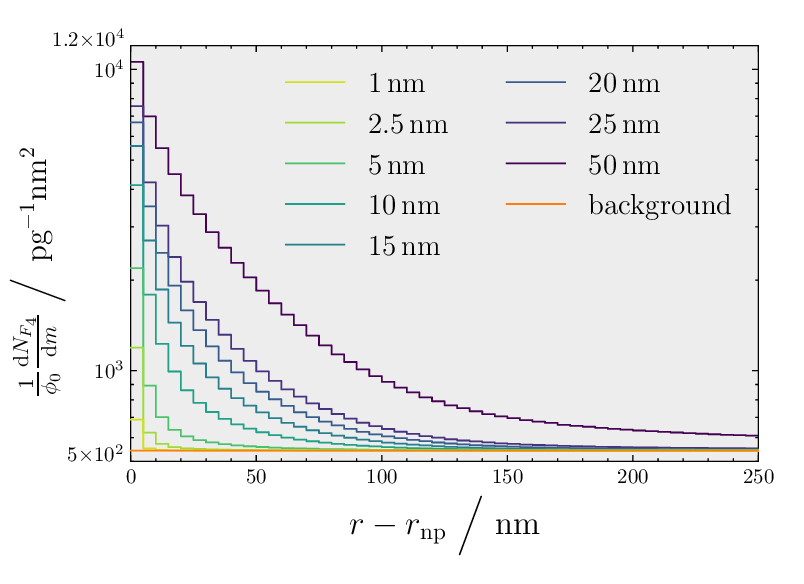}
				\end{center}
			\end{minipage}
			\caption{
				\textbf{\protect\subref{fig:ID_all}} Variation of the radial density
				of ionization clusters of size \mbox{$\nu\ge 4$} $\dd N_{F_4} / \dd r$
				normalized to primary fluence $\phi_0$ as a function of distance from
				nanoparticle surface \mbox{$r - \rnp$}.
				\textbf{\protect\subref{fig:g_all}} Variation of the ionization cluster
				dose $\dd N_{F_4} / \dd m$, again for clusters of size \mbox{$\nu\ge 4$}
				normalized to the primary fluence $\phi_0$, as a function of distance from
				the nanoparticle surface \mbox{$r - \rnp$}.
			}
			\label{fig:result_ID_g_all}
		\end{minipage}
	}
\end{figure}

Fig.~\subref*{fig:ID_all} illustrates the frequency distribution of ionization
clusters of size \mbox{$\nu\ge 4$} within shells surrounding the nanoparticles $\dd
	N_{F_4}/\dd r$ normalized to the primary fluence $\phi_0$. The count of clusters has
been corrected for contributions that occur when the gold NP is replaced by water,
as outlined in Eq. \ref{eq:background-correction}. In alignment with the
$\rnp^3$-dependence discussed in Section~\ref{subsection:fluence_spectra}, the
nanoparticle radius $\rnp$ impacts the total number of clusters created. The
background contribution $\dd N_{F_4}^\mathrm{b}/\dd r$ is proportional to the volume
comprised between $[r, r+ \dd r)$, which approximately grows with $r^2$.

\begin{figure}[t!]
	\centering
	\fbox{
		\begin{minipage}{0.667\textwidth}
			\begin{center}
				\includegraphics[width=0.75\textwidth]{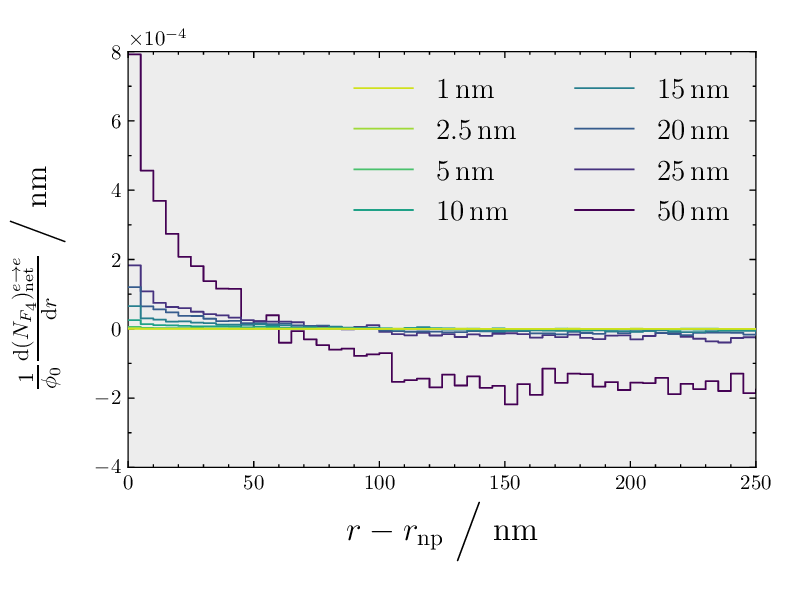}
			\end{center}
			\vspace{-5mm}
			\caption{Distribution of the net number of
				ionization clusters $\dd (N_{F_4})_\mathrm{net}^{e\to e} / \dd r$ normalized to
				primary fluence $\phi_0$ as a function of distance from the nanoparticle surface
				\mbox{$r - \rnp$}, i.e. without the background-correction, as defined
				in Eq.~\ref{eq:net-contribution}.
			}
			\label{fig:result_ID_diff_ee}
		\end{minipage}
	}
\end{figure}

Fig.~\ref{fig:result_ID_diff_ee} shows the net effect of the gold nanoparticle of
clusters stemming from electron impact ionizations, created by electrons entering
the gold nanoparticle, $\dd (N_{F_4})_\mathrm{net}^{e\to e} / \dd r$ normalized to
primary fluence $\phi_0$. While the attenuation coefficients for photons differ by
two or three orders of magnitude between water and gold at the photon energies of
the 100 kVp spectrum used here, the stopping powers of electrons at the relevant
energies are of the same order of magnitude\footnote{\ The fluence-averaged
	stopping-power ratio is estimated to be $\braket{S_\mathrm{w}}_{\phi_1^e} /
		\braket{S_\mathrm{g}}_{\phi_1^e} \approx 2.48$ \citep{ICRU_37}}. Therefore, the gold
and water-only contributions of ionization clusters almost cancel each other out. In
fact, beyond distances of $60\,\nm$ - $100\,\nm$, the water-only contribution exceeds
the gold contribution, which is known for energy imparted as the sink effect
\citep{Brivio_2017}.

Since the absolute number of clusters from the background grows with shell volume,
Fig.~\subref*{fig:g_all} displays cluster frequency per mass $\dd m$, of the
corresponding shell. In this normalization, the background contribution to the
resulting quantity is constant in $r$ and the resulting quantity corresponds to the
cluster dose $g^{F_4}$ introduced by \mbox{\citep{Faddegon_2023}}, conceptually.

\begin{figure}[t!]
	\fbox{
		\begin{minipage}{\textwidth - 4mm}
			\begin{minipage}{.5\textwidth}
				\subfloat[]{\raggedright{}\label{fig:dNdr_C-1}}\vspace{-0.35cm}
				\begin{center}
					\includegraphics[width=\textwidth]{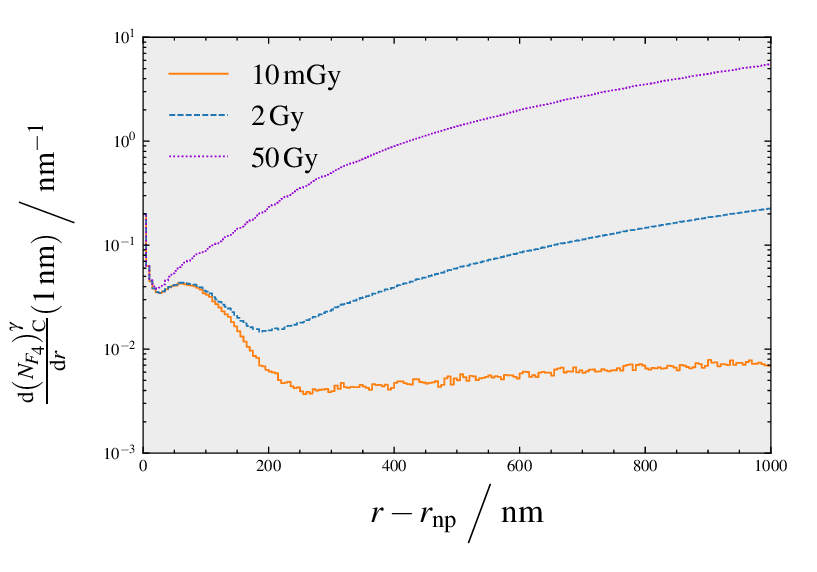}
				\end{center}
			\end{minipage}%
			\begin{minipage}{.5\textwidth}
				\subfloat[]{\raggedright{}\label{fig:dNdr_C-2}}\vspace{-0.35cm}
				\begin{center}
					\includegraphics[width=\textwidth]{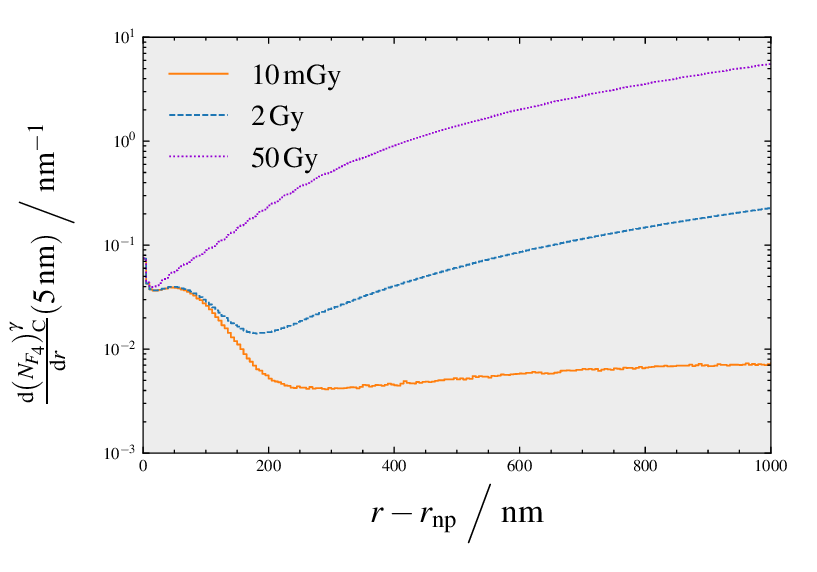}
				\end{center}
			\end{minipage}
			\vspace{-0.7mm}
			\begin{minipage}{.5\textwidth}
				\subfloat[]{\raggedright{}\label{fig:dNdr_C-3}}\vspace{-0.35cm}
				\begin{center}
					\includegraphics[width=\textwidth]{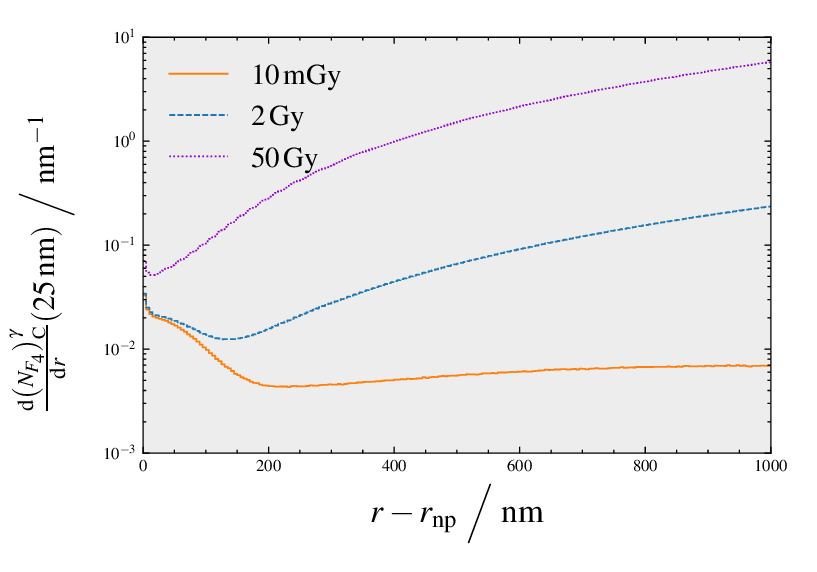}
				\end{center}
			\end{minipage}%
			\begin{minipage}{.5\textwidth}
				\subfloat[]{\raggedright{}\label{fig:dNdr_C-4}}\vspace{-0.35cm}
				\begin{center}
					\includegraphics[width=\textwidth]{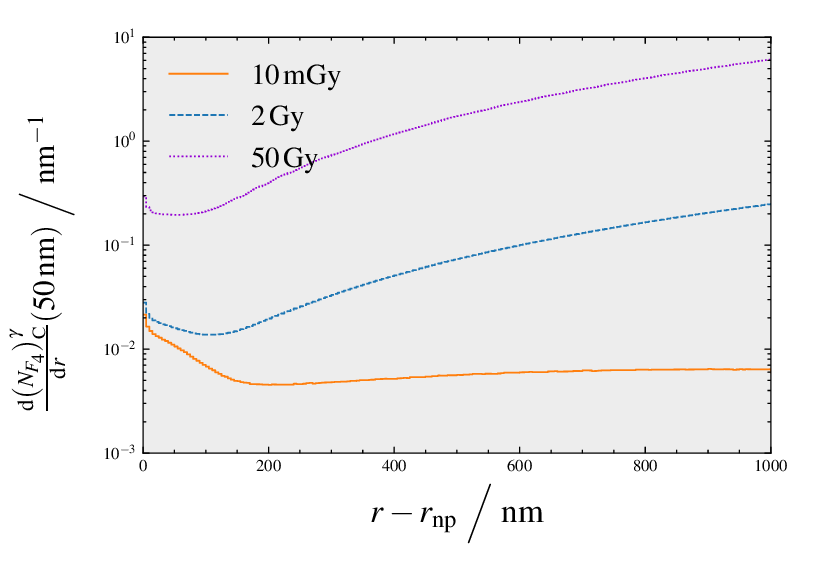}
				\end{center}
			\end{minipage}
			\caption{
				Conditional radial density of ionization clusters (cf. Eq.~\ref{eq:dQdx_C}) for different
				chosen dose values: $10\,\mGy$ (orange), $2\,\Gy$ (blue) as well as $50\,\Gy$ (purple)
				for the different nanoparticle radii
				\textbf{\protect\subref{fig:dNdr_C-1}}	$\rnp = 1\,\nm$,
				\textbf{\protect\subref{fig:dNdr_C-2}}	$\rnp = 5\,\nm$,
				\textbf{\protect\subref{fig:dNdr_C-3}}	$\rnp = 25\,\nm$ and
				\textbf{\protect\subref{fig:dNdr_C-4}}	$\rnp = 50\,\nm$.
			}
			\label{fig:dNdr_C}
		\end{minipage}
	}
\end{figure}

The radial cluster density of photons conditional on the occurrence of at least one
ionization as described by Eq.~\ref{eq:dQdx_C} is depicted in
Figs.~\subref*{fig:dNdr_C-1}~-~\subref*{fig:dNdr_C-4} for different radii ($\rnp =
	1\,\nm$, $5\,\nm$, $25\,\nm$ and $50\,\nm$) as well as three doses corresponding to
different use cases: $10\,\mGy$ as a typical order of magnitude of the dose
received in a CT scan, $2\,\Gy$ as a typical fraction in conventional radiation
therapy as well as $50\,\Gy$, corresponding to a possible single dose fraction in
FLASH-RT that utilizes brief ($<200\,\mathrm{ms}$) irradiations with dose rates $\ge
	40\,\Gy/\mathrm{s}$ \citep{Vozenin_2019}.

It becomes apparent, that the absorbed dose influences the formation of clusters in a
non-linear fashion, which is a consequence of the likelihood of multiple interactions
occurring, which increases with the absorbed dose, outlined in
Section~\ref{subsection:normalization}. This effect is especially evident for
smaller gold NPs, where a shoulder pattern emerges for distances of roughly up to
$200\,\nm$ from the nanoparticle surface. Judging from this distance, these clusters
are caused by low energy electrons, namely the M shell Auger electrons visible in
the photon-generated electron spectrum in Fig.~\ref{fig:surface_fluence_ge}.

\subsection{Imparted energy and dose}
\label{subsection:eimp}

\begin{figure}[t!]
	\fbox{
		\begin{minipage}{\textwidth - 4mm}
			\begin{minipage}{.5\textwidth}
				\subfloat[]{\raggedright{}\label{fig:eimp_all}}\vspace{-0.35cm}
				\begin{center}
					\includegraphics[width=\textwidth]{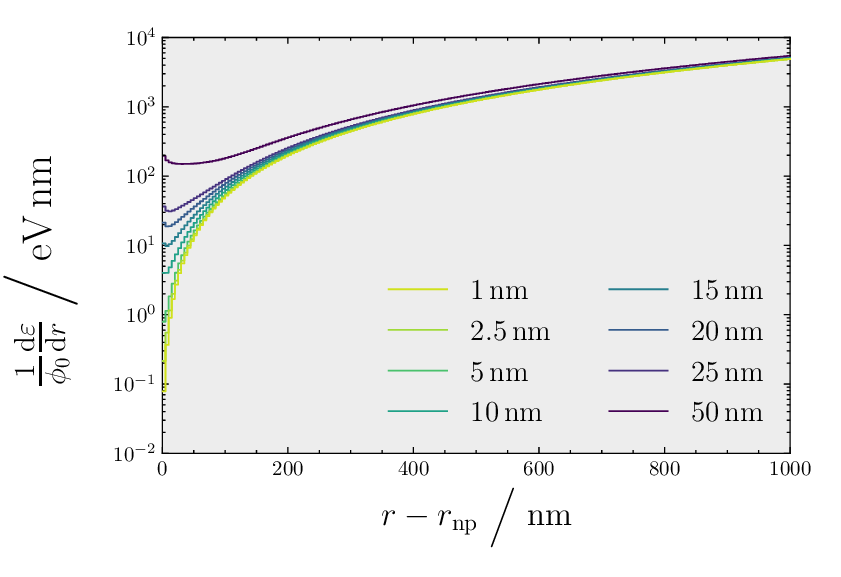}
				\end{center}
			\end{minipage}%
			\begin{minipage}{.5\textwidth}
				\subfloat[]{\raggedright{}\label{fig:dose_all}}\vspace{-0.35cm}
				\begin{center}
					\includegraphics[width=\textwidth]{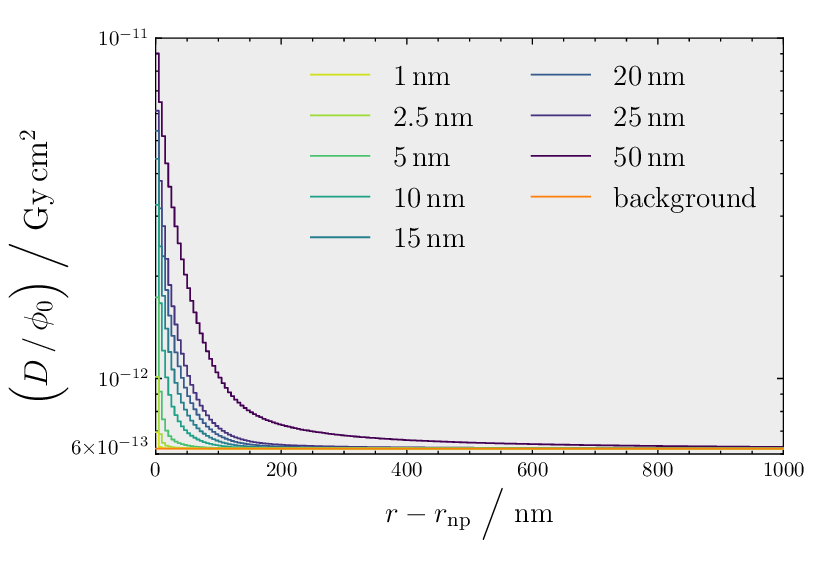}
				\end{center}
			\end{minipage}
			\caption{\textbf{\protect\subref{fig:eimp_all}} Distribution of total energy
				imparted per radial shell \mbox{$\dd \varepsilon / \dd r$} normalized to
				primary fluence $\phi_0$ as a function of distance from the nanoparticle surface
				\mbox{$r - \rnp$}.
				\textbf{\protect\subref{fig:dose_all}} Distribution of absorbed dose
				contribution $D = \dd \varepsilon / \dd m$ from the gold nanoparticle normalized to primary
				fluence $\phi_0$ as a function of distance from the nanoparticle surface
				\mbox{$r-\rnp$}.
			}
			\label{fig:result_eimp_all}
		\end{minipage}
	}
\end{figure}

The imparted energy per radial shell, illustrated in Fig.~\subref*{fig:eimp_all}, as
well as the absorbed dose (Fig.~\subref*{fig:dose_all}) per primary fluence $\phi_0$
as a function of radial distance from the nanoparticle surface qualitatively
correspond to the ionization cluster density and cluster dose
(Figs.~\subref*{fig:ID_all}~and~\subref{fig:g_all}, respectively) and display the
same $\rnp^3$-dependence discussed earlier.

\begin{figure}[t!]
	\centering
	\fbox{
		\begin{minipage}{0.667\textwidth}
			\begin{center}
				\includegraphics[width=0.75
					\textwidth]{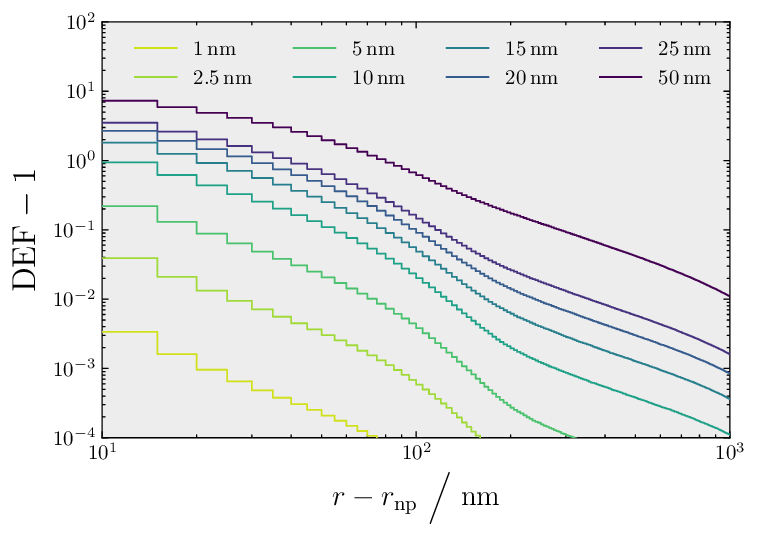}
			\end{center}
			\vspace{-5mm}
			\caption{Deviation from unity of the dose enhancement factor as a
				function of distance to the nanoparticle \mbox{$r - \rnp$} for the different
				nanoparticle radii. The radii \mbox{$\rnp = 25\,\nm$} and \mbox{$\rnp =
						50\,\nm$} are compared to the data of \citep{Rabus_Li_2021}, in
				Supplementary
				Figs.~\protect\subref*{fig:DEF-25-comparison}~and~\protect\subref*{fig:DEF-50-comparison},
				respectively.
			}
			\label{fig:DEF-1}
		\end{minipage}
	}
\end{figure}

The dose enhancement factor (DEF, also referred to as dose enhancement ratio, DER),
defined as the ratio of absorbed dose in a volume surrounding a gold NP to the
corresponding absorbed dose in the absence of the gold NP in the same volume, is
commonly used to characterize the dosimetric impact of nanoparticles, especially in
simulation studies. The EURADOS intercomparison study \citep{Rabus_Li_2021} reports
the enhancement as ``deviation of the DER from 1'', $\mathrm{DEF}-1$. For comparison,
this quantity is displayed in Fig.~\ref{fig:DEF-1}. A detailed comparison of the
DEFs for nanoparticles of \mbox{$\rnp = 25\,\nm$} and \mbox{$\rnp = 50\,\nm$} can be
found in Supplementary
Figs.~\subref*{fig:DEF-25-comparison}~and~\subref*{fig:DEF-50-comparison},
respectively\footnote{\ Note that the nanoparticle sizes in \cite{Rabus_Li_2021} are
	specified as diameters, i.e. a nanoparticle with radius $\rnp = 50\,\nm$ is referred
	to as a ``$100\,\nm$-nanoparticle''.}.

\subsection{Linking ionization clustering to imparted energy and dose}
\label{subsection:link}

In order to relate ionization clustering to the absorbed dose, a cluster enhancement
factor (CEF$_{N_{F_k}}$) is defined, analogously to the dose enhancement factor
(DEF), as the ratio of the frequency of clusters of size \mbox{$\nu\ge k$} in a
volume surrounding a gold NP (or shells around it) to the corresponding frequency in the
absence of the gold NP.

\begin{figure}[t!]
	\fbox{
		\begin{minipage}{\textwidth - 4mm}
			\centering
			\begin{minipage}{.5\textwidth}
				\vspace{-2mm}
				\subfloat[]{\raggedright{}\label{fig:diff}}\vspace{-0.35cm}
				\begin{center}
					\vspace{3mm}
					\includegraphics[width=0.92\textwidth]{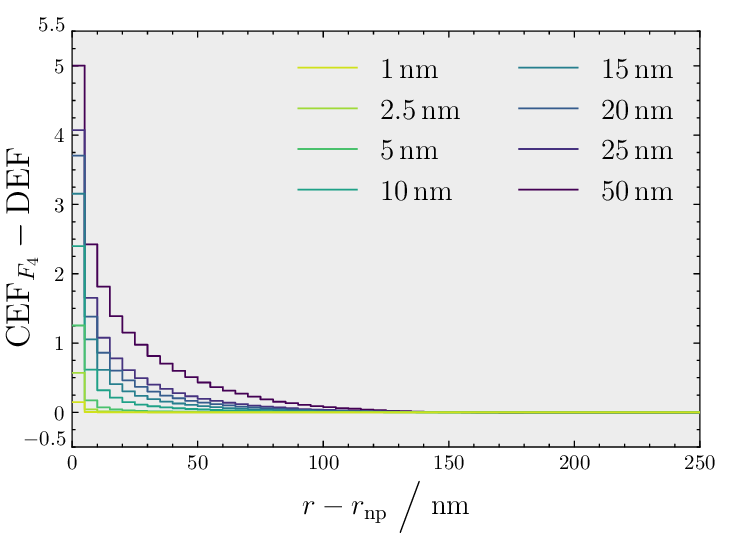}
				\end{center}
			\end{minipage}%
			\begin{minipage}{.5\textwidth}
				\subfloat[]{\raggedright{}\label{fig:NFk_per_eimp}}\vspace{-0.35cm}
				\begin{center}
					\includegraphics[width=\textwidth]{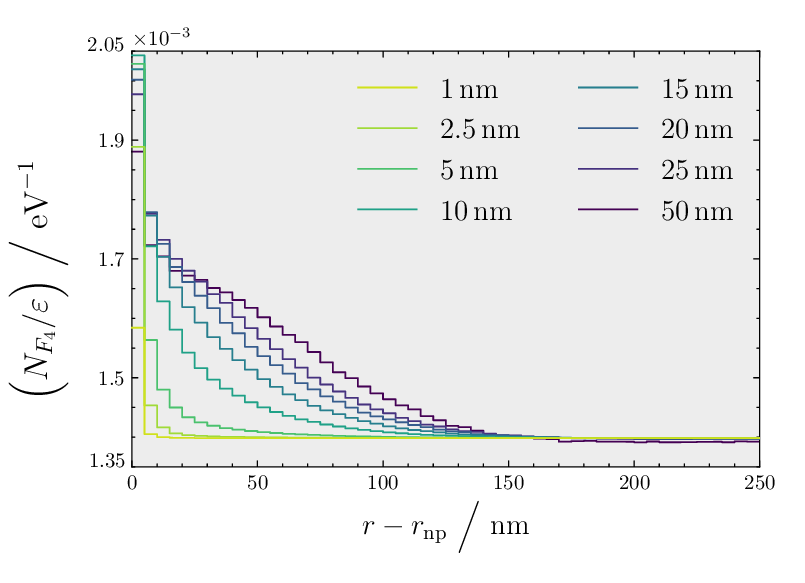}
				\end{center}
			\end{minipage}
			\vspace{-5mm}
			\caption{
				\textbf{\protect\subref{fig:diff}} The difference of the cluster
				enhancement factor (CEF$_{N_{F_4}}$) and dose enhancement factor (DEF) within the
				first $250\,\nm$ from the nanoparticle surface. The two enhancement
				factors can be found in Supplementary
				Figs.~\protect\subref*{fig:CEF}~and~\protect\subref*{fig:DEF}.
				\textbf{\protect\subref{fig:NFk_per_eimp}} Ratio of frequency of
				clusters of size \mbox{$\nu\ge 4$} per shell and energy imparted in that shell.
			}
			\label{fig:DEFCEF}
		\end{minipage}
	}
\end{figure}

The difference between CEF$_{N_{F_k}}$ and DEF is reported in Fig.~\subref*{fig:diff} and
illustrates a relatively increased enhancement of clustering with respect to the
enhancement of dose in the shells surrounding the gold NP. This effect is highly
localized and largely confined to radial distances up to $\lesssim 50\,\nm$, albeit
quite pronounced. The enhancement factor of ionization clustering as well as the
dose enhancement factor, for comparison, are reported individually in Supplementary
Figs.~\subref*{fig:CEF}~and~\subref*{fig:DEF}, respectively.

Fig.~\subref*{fig:NFk_per_eimp} shows the number of clusters per energy imparted as
a function of radial distance from the nanoparticle. Both of these results are
consistent with the emission of Auger electrons from the nanoparticle: the energy
imparted by these low energy Auger electrons results from a higher number of
ionizations.

\newpage
\section{Discussion}
\label{section:discussion}

\subsubsection*{Fluence spectra}

The photon (and consequentially the electron fluence) impinging on the nanoparticle
is influenced by photons scattered back into the region of interest. The extent of
this effect depends on the geometry considered. The geometry here has been designed
to make the results comparable to the \mbox{EURADOS} intercomparison exercise
\citep{Li_2020a,Li_2020b,Rabus_Li_2021,Rabus_2021}, as outlined in
Section~\ref{subsection:simulation}. Here, the total number of photons present in
the region of interest is roughly $1.38$ times that of primary photons.

Profiting from the two-step simulation scheme, fluence spectra of electrons could be
individuated for both photons (Fig.~\ref{fig:surface_fluence_ge}) and electrons
(Fig.~\subref*{fig:fluence_ee} and Fig.~\subref*{fig:fluence_ee_log}) impinging on
the nanoparticle surface. Whereas the electrons produced by electrons are largely
independent of NP radius, the photon-produced electron fluence grows roughly
linearly with the NP radius and it could be shown that with growing NP-radius, it is
electrons produced by incident photons that dominate the fluence of electrons
leaving the nanoparticle. The peaks in Fig.~\ref{fig:surface_fluence_ge} have been
identified as electrons stemming from Auger transitions. The small range of these
electrons is assumed to drive the increase in clustering in the proximity to the NP
surface.

The spectral energy fluence of such photon-induced electrons aligns well with the
results from the EURADOS intercomparison publication by Li et al. (2024, in press),
which has investigated nanoparticles of $\rnp = 25\,\nm$ and $\rnp = 50\,\nm$.
Comparisons of the spectra are shown in Supplementary
Figs.~\subref*{fig:fluence-ge-25-comparison}~and~\subref*{fig:fluence-ge-50-comparison}
for these two NP radii, respectively.

\subsubsection*{Clustering methods}

The associated volume clustering (AVC) approach, initially used to score clusters of
energy transfer points, has been used here for the first time to score ionization
clusters to assess the physical radiation effect of NPs and at target size
equivalent to those considered, for instance, in \cite{Grosswendt_2006} and
\cite{Faddegon_2023}. Broadly speaking, this method and other sampling-based methods
\citep{Alexander_2015, Ramos-Mendez_2018, Selva_2018} build one of three categories
of clustering methods, all of which realize the ten basepair localization criterion
mentioned in Section~\ref{subsection:clustering} in one way or another. The other
two categories are 1.) counting ionization-sites within some type of grid
\citep{Braunroth_2020, Rabus_2020, Ngcezu_2021} and 2.) density-based clustering
such as the Density-Based Spatial Clustering of Applications with Noise (DBSCAN)
algorithm, initially developed by \cite{Ester_1996} and modified for the estimation
of DNA lesion clusters estimated as a random subset of ionizations
\citep{Francis_2011}.

Unlike uniform sampling, sampling in the AV of a track utilizes each ionization for
scoring and thus makes more efficient use of sampled volumes. How much more
efficient this is depends on the fraction of the AV and the scoring volume.

While intended use of DBSCAN is not the clustering of ionizations but rather of the
(more sparsely distributed) DNA lesions, one of its pathologies is its tendency to
form clusters of large spatial dimensions, as it is based on a spatial, pairwise
distance measure. Such complex damages extended over volumes larger than a ten
basepair DNA segment may comprise many ionizations and their significance to DNA
damage is difficult to interpret. The associated volume clustering method avoids
this behaviour by design, strictly enforcing a spatial limit to the size of a
cluster.

Grid-based approaches\citep{Braunroth_2020, Rabus_2020}, e.g. in cylindrical grids,
are computationally efficient (with a time complexity of $\mathcal{O}(\bar N)$, with
the expected number $\bar N$ of ionization sites of a track). However, AVC, while
not solvable in linear time ($\mathcal{O}(\bar N^2)$\footnote{\ While this is not a
	challenge when clustering ionizations from $100\,\mathrm{kVp}$ X-rays, this might be
	a relevant consideration for high-LET radiation.}), can be utilized with virtually
any scoring volume geometry, such as spheres used here or cylinders, which more
closely resemble the shape of a DNA segment.

\subsubsection*{Main findings}

Different contributions to ionization clusters were quantified. Together with the
scored fluence spectra on the nanoparticle surface, these contributions could reveal
information about the physical mechanisms characterizing the gold NP enhancement
effect. With growing NP-radius, it is electrons produced by incident photons that
dominate the fluence of electrons leaving the nanoparticle and thus the occurrence
of ionization clusters.

For larger distances from the surface, the linear density of ionization
cluster-frequency (Fig.~\subref*{fig:ID_all}) is dominated by the background
contribution. While itself constant throughout space, this contribution grows with
the volume of the shell around the nanoparticle (which is roughly proportional to
$r^2$).

A straightforward way to visualize the range of significant increase of ionization
clusters is to consider the number of ionization clusters per mass,
Fig.~\subref*{fig:g_all}. In this representation, the background contribution is
constant. In reminiscence of the energy imparted per mass -- the dose --
\cite{Faddegon_2023} have coined the term `cluster dose' for this quantity.

As worked out in Section~\ref{subsection:normalization}, the primary fluence affects
the formation of ionization clusters. The plots of the conditional radial density of
clusters (Figs.~\subref*{fig:dNdr_C-1}~-~\subref*{fig:dNdr_C-4}) show the
non-linearity introduced by multiple occurrences of photon interactions at different
dose levels.

This emphasizes the necessity of considering absolute (i.e. fluence dependent)
quantities rather than relative quantities, like enhancement ratios or cluster
frequencies per imparted energy, alone. This effect is especially evident for
smaller gold NPs, where a shoulder pattern emerges for distances fitting the range
of the M shell Auger electrons visible in the photon-generated electron spectrum in
Fig.~\ref{fig:surface_fluence_ge}.

The results for the energy imparted within the shells around the NPs have been used
to calculate absorbed dose as well as dose enhancement factors (DEFs). These DEFs
show good agreement with the results from the EURADOS intercomparison
\citep{Rabus_Li_2021} exercise, which have been used to check the results for
consistency, see Supplementary
Figs.~\subref*{fig:DEF-25-comparison}~and~\subref*{fig:DEF-50-comparison}, for DEFs
for nanoparticle radii $25\,\nm$ and $50\,\nm$, respectively. It should be noted
that only energy imparted by ionizations was scored, whereas the in intercomparison
exercise ionizations as well as excitations were scored. This contribution
has been found to be $80\,\%$ of all energy deposits \citep{Gervais_2006}. Another
difference worth mentioning is the two-step simulation setup. Simulations conducted
within the intercomparison exercise were performed with a circular beam of
\mbox{$d_\mathrm{np} + 10\,\nm$}, with the nanoparticle diameter $d_\mathrm{np}$.
The resulting lack of secondary electron equilibrium was accounted for using the
methods detailed in \cite{Rabus_2019}.

While the occurrence of ionization clusters (Fig.~\subref*{fig:ID_all}) and the
energy imparted (Fig.~\subref*{fig:eimp_all}) show qualitative similarities, it
could be shown that the number of ionization clusters of size \mbox{$\nu\ge 4$} per
imparted energy (Fig.~\subref*{fig:NFk_per_eimp}) varies with distance and is
increased within the first $150\,\nm$ from the nanoparticle surface. This enhanced
clustering near the NP surface may be the origin of what is described as
therapy-activated nanoparticle and immunotherapy \citep{Penninckx_2023}. In
alignment with this local increase of clustering, the determined cluster enhancement
factor (CEF$_{F_4}$) exceeds the DEF by up to $46\,\%$ (for the $10\,\nm$-radius
gold NP) (Fig.~\subref*{fig:diff} as well as individually in Supplementary
Figs.~\subref*{fig:CEF}~and~\subref*{fig:DEF}). This highlights the necessity of
considering ionization clustering for the characterization of cell damage.

\subsubsection*{Relevance}

Given that in a therapeutic application of gold NPs there is generally a large
number of gold NPs present in a tumour cell, the question of the relevance of the
present results may be raised. To address this point, the following estimates are
based on the approach outlined in \citep{Rabus_2024b}. Assume a spatially uniform
distribution of gold NPs and a mass fraction of $2\,\%$ of gold in a cell
represented by water. For the photon energy spectrum considered in this work, such a
concentration of gold results in an increase of the average absorbed dose by a
factor of about $1.8$. The corresponding volume fraction of gold NPs is about
$10^{-3}$ such that the `associated volume' of a gold NP (comprising the points in
space closer to the gold NP than to its neighbours) has a radius of about ten times
the gold NP radius. The nearest neighbour distance is thus about $20$ times the gold
NP radius. This implies that for this value of mass fraction of gold, the nearest
neighbours are outside the region of significant enhancement of ionization
clustering (or energy imparted) for the gold NP radii considered here, as can be
seen in Fig.~\subref*{fig:g_all}. For uniformly distributed nanoparticles of these
sizes, the density of ionization clusters (or energy imparted) can be assumed to be
additive. For smaller gold NP radii, the nearest neighbours are within the region of
significant increase of ionization clustering and can interact with the electrons
emitted from the considered gold NP where a photon interaction occurs.

When the nearest neighbors are located outside of the region of significant
clustering there is only a small modification of the spatial density of ionization
clusters in the form of a small sink effect that has been observed for larger
nanoparticles at distances exceeding $100\,\nm$ (Fig.~\ref{fig:result_ID_diff_ee})
of the neighbouring gold NPs and a localized increase around them (this estimation
also holds for the number of ionization clusters per imparted energy,
Fig.~\subref*{fig:NFk_per_eimp}). Hence, the synergistic effects of the gold NPs can
also be expected to be additive as well, and this conjecture will be tested in
future work. It should be added that according to the review by
\cite{Kuncic_Lacombe_2018}, the maximum concentrations of gold NPs found in
radiobiological studies are more than two orders of magnitude smaller than the mass
fraction of $2\,\%$ used in the estimate above. With such concentrations of gold
NPs, the nearest neighbour distance for uniform spatial distribution is higher by a
factor in the order of 50. Then, neighbours of gold NPs of the smallest radius of
$1\,\nm$ are also outside the range of significantly increased ionization cluster
density. When clustering of nanoparticles occurs (as is often found in
radiobiological studies), the preceding argument does not hold and changes due to
the close proximity of nearest neighbours will occur as reported in preceding
simulation studies \citep{Byrne_2018,Rudek_2019}. However, it may still be possible
to estimate the respective overall effect by superposition of the effects of a
single gold NP reported here, which will also be investigated in further studies.

It is worth mentioning that the presence of further gold NPs in a considered volume
of matter not only changes the radiation field of electrons but also that of
photons. In the energy range considered in this study, this will be almost
exclusively through coherent photon scattering, which contributes on average about
$6\,\%$ of the total interaction coefficient of gold. In addition, there is a small
contribution to the photon fluence by fluorescence photons originating in the
radiative de-excitation of gold atoms undergoing a core-level
photoabsorption\footnote{\ In this context, it may be useful to comment on the
	absence of K shell fluorescence peaks in the photon fluence spectra induced by
	incident photons (Supplementary
	Figs.~\subref*{fig:surface_fluence_gg}~and~\subref{fig:surface_fluence_gg_diff}). As
	can be derived from the evaluated photon data library \citep{Cullen_1997}, the
	probability of a K shell photoabsorption occurring in gold can be estimated to be
	about $73\,\%$ in the energy range above the K absorption edge of gold for the
	photon spectrum used here. According to the evaluated atomic data library
	\citep{EADL}, the probability of a radiative deexcitation of a K shell vacancy in
	gold is $95\,\%$, where the dominant K$\alpha$1 line occurs with a probability of
	about $48\,\%$. Therefore, one might expect to see peaks related to fluorescence
	decay of K shell ionized gold atoms. The main reason for their absence is the use of
	an energy bin size of $100\,\eV$ in the scoring, which significantly reduces the
	height of any, generally very narrow peaks. Furthermore, averaged over the complete
	incident photon spectrum, the proportion of K shell absorption to the total
	photoabsorption cross-section is only about $1.8\,\%$. Therefore, the probability of
	a gold K$\alpha$1 photon at about $68.805\,\keV$ \citep{EADL} being emitted is only
	about $0.9\,\%$. Accordingly, the content of the corresponding photon energy bin is
	only increased by this small proportion, which is well below the statistical
	fluctuations.}.

\subsubsection*{Caveats and final remarks}

The two-step simulation approach entails sampling photons and electrons on the
surface of a gold nanoparticle. The assumption made here is that the particles
impinging on the nanoparticle are distributed isotropically, which is an
approximation. Subsequently cluster frequencies (or other quantities) are considered
solely as a function of radial distance, overlooking anisotropies of the
distribution of ionization clusters. The anisotropy of photoelectron emission has
been studied \citep{Derrien_2023,Rabus_2024}. The magnitude of the anisotropy of
imparted energy under CPE is estimated to be in the few percent range ($3.5\,\%$ for
$30\,\keV$ photons and $3.9\,\%$ for $80\,\keV$ photons) for a $25\,\nm$-radius
target at a distance of $200\,\nm$ from the surface \citep{Rabus_2024}.

The sampling method employed for single incident electrons is grounded on the
assumption that these electrons are independent. Of course, this is an approximation
that discards synergistic effects of electrons with other electrons from the same
primary photon. If the interactions of complete electron tracks with gold NPs were
to be considered, their ``fluence'' would be much smaller than the fluence
determined for individual electrons. Therefore, a determination of conditional
quantities as undertaken for photons would be much more intricate.

While the geometry used in this work has been designed to ensure CPE conditions,
photon equilibrium conditions are only met partially: the geometry allows for
scoring photons that are scattered back into the scoring volume (region
`$\mathbf{A}$' in Fig.~\ref{fig:simulation-setup}), although photons scattered at
larger angles e.g. laterally into the scoring volume are not accounted for as the
chosen photon beam is collimated to the (extended) region of interest (i.e. regions
`$\mathbf{A}$' and `$\mathbf{B}$' in Fig.~\ref{fig:simulation-setup}). It is assumed
that this is a small effect \citep{Rabus_2019} that should be addressed in a future
study.

Conditional ionization cluster densities have been determined for photons using the
estimates of the mean expected number of ionization interactions for photons that
may cause inner-shell ionizations. The probability of an ionizing interaction is
considerably larger, when considering low energy ($E\le1\,\keV$) photons. However,
the fluence of these photons is negligibly small and these photons are not capable
of producing M shell Auger electrons so that their impact is expected to be small.

The simulations in this work have been conducted for gold in water. Gold
nanoparticles that have been considered for clinical usage are coated to influence
cell uptake and mitigate toxic effects. The effect that such coatings can have on
the number of resulting ionization clusters has not been studied here, but will
influence low-energy electrons
\citep{Rabus_2023,Morozov_2018,Belousov_2018,Belousov_2019}. The size of
nanoparticles used in experiments is also subject to variation due to manufacturing,
a source of uncertainty that could be estimated for comparison with experimental
data. Finally, although track-structure codes, in principle, allow for highly
accurate computations, the uncertainties in the cross-sections used for simulation
are assumed to be a major source of uncertainty, most noticeably at low energies
\citep{Bug_2014,Villagrasa_2018,Villagrasa_2022}. Especially when investigating
clustering caused by such low energy electrons in the proximity of a nanoparticle,
the effect of such errors should be gauged.

\subsubsection*{Conclusions}

The simulations in this work have been conducted under charged particle equilibrium
conditions as well as using a track structure code for simulation. While previous
works have examined the radial dependence of energy imparted around a gold
nanoparticle, this work has also determined the frequency of ionization clusters.
This methodology as well as the presented validation of the data allow for a
thorough systematic evaluation of the gold nanoparticle enhancement effect. The fact
that calculations have been conducted under charged particle equilibrium conditions
allows for a realistic evaluation of the cluster and dose enhancement factors and
the quantitative comparison of the magnitude of these effects.

In fact, the data presented show that the gold NPs enhancement of ionization cluster
frequency exceeds that of absorbed dose in the immediate vicinity of the gold NP so
that the energy imparted cannot act as a good proxy for increased ionization cluster
frequencies and highlights the necessity for nanodosimetric evaluation to understand
physical mechanisms behind their radiation effect. The reported conditional radial
densities of ionization clustering show a shoulder increasing for smaller
nanoparticle radii that -- judging from the distance to the surface -- can be
explained as contributions of the M shell Auger electrons.

The two-step approach in simulation allowed for the individual consideration of the
spectral fluences of electrons produced by photons and electrons. For the
$100\,\mathrm{kVp}$ X-rays studied in this work, the relevant component to
clustering has been shown to be the photon field-induced electrons, rather than the
electron field itself, which -- apart from a small sink effect -- is not
significantly affected by the gold nanoparticles. A rather important role seems to
be the interplay between the production of electrons within the nanoparticle and
their capability of leaving it. Nevertheless the electron field is important in
determining the background of ionization clusters as well as energy imparted.

As an important driver for both the local increase of ionization clusters per energy
imparted and the local increase of the conditional radial densities of ionization
clustering for smaller nanoparticle radii, these are found to be mediated by an
increase in low-energy electrons consistent with the emission of electrons from
M shell Auger transitions as part of the de-excitation cascade from
inner-shell-ionized gold atoms.

\newpage
\section*{Acknowledgements}
\label{section:acknowledgements}

The authors express their gratitude to the dedicated team of the High Performance
Cluster of the German National Metrology Institute (PTB) for their ongoing support
throughout the production of the data.

This project is part of the programme ``Metrology for Artificial Intelligence in
Medicine'' (M4AIM), which is funded by the Federal Ministry for Economic Affairs and
Climate Action (BMWK) within the scope of the ``QI-Digital'' initiative.

\footnotesize

\bibliographystyle{apalike}
\bibliography{references}

\newpage
\appendix
\setcounter{page}{1}
\setcounter{figure}{0}
\setcounter{table}{0}

\renewcommand{\thefigure}{S\arabic{figure}}
\captionsetup[figure]{labelformat=supplementaryfigure}
\renewcommand{\thetable}{S\arabic{table}}
\captionsetup[table]{labelformat=supplementarytable}

\section*{Supplementary Material}
\label{section:supplemenary}

\vspace{1 cm}
\subsubsection*{Inelastic interaction probability in gold}
\begin{figure}[H]
	\centering
	\fbox{
		\begin{minipage}{.67\textwidth}
			\begin{center}
				\includegraphics[width=0.75\textwidth]{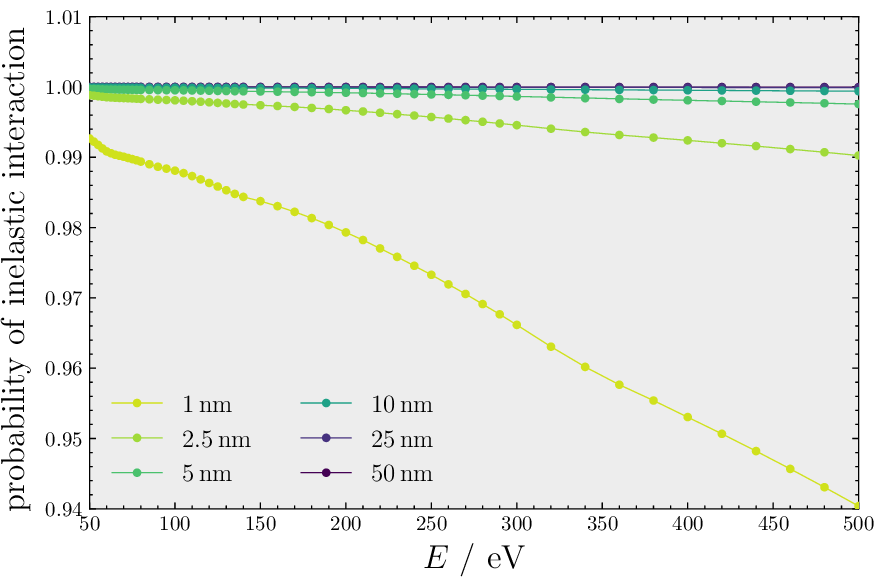}
			\end{center}
			\vspace{-5mm}
			\caption{Estimated probability of an incident electron in the energy range
				between $50\,\eV$ and $500\,\eV$ undergoing an inelastic interaction in gold NPs of
				different radii (see legend). The probability was estimated using the
				electron mean free paths extracted from the database of the 2018 release of
				Penelope \citep{Penelope} deploying the \texttt{tables.f} program provided with
				the code package. The estimate is based on the simplifying assumption that elastic
				scattering is only in forward direction.}
			\label{fig:p_penelope}
		\end{minipage}
	}
\end{figure}

\subsubsection*{Probability of photon interactions in a gold NP}
\begin{table}[H]
	{\footnotesize
		\begin{minipage}{\textwidth}
			\captionsetup{justification=raggedright,
				singlelinecheck=false
			}
			\caption{Estimates of the probability of a photon
				interaction taking place calculated via the linear
				attenuation coefficient as detailed in Eq.~\ref{eq:p_g} as well as
				over the loss of photons within the gold NP obtained from the
				simulation as detailed in Eq.~\ref{eq:p_g_sim}. The fluence
				of photons leaving the nanoparticle used in this estimate is depicted
				in Supplementary Fig.~\protect\subref*{fig:surface_fluence_gg_diff}.
			}
			\begin{center}
				\def\arraystretch{1.5}
				\begin{tabular}{l|c|c|c}
					$\rnp$     & $\left<p_i^\gamma\right>$ (Eq.~\ref{eq:p_g}) & $\left<p_i^\gamma\right>$ (Eq.~\ref{eq:p_g_sim}) & rel. diff. \\ \hline\hline
					$1\,\nm$   & $3.02\smalltimes 10^{-5}$                    & $2.94\smalltimes 10^{-5}$                        & 2.94\,\%   \\ \hline
					$2.5\,\nm$ & $7.56\smalltimes 10^{-5}$                    & $7.36\smalltimes 10^{-5}$                        & 2.75\,\%   \\ \hline
					$5\,\nm$   & $1.51\smalltimes 10^{-4}$                    & $1.48\smalltimes 10^{-4}$                        & 2.33\,\%   \\ \hline
					$10\,\nm$  & $3.02\smalltimes 10^{-4}$                    & $2.93\smalltimes 10^{-4}$                        & 3.14\,\%   \\ \hline
					$15\,\nm$  & $4.54\smalltimes 10^{-4}$                    & $4.41\smalltimes 10^{-4}$                        & 2.86\,\%   \\ \hline
					$20\,\nm$  & $6.05\smalltimes 10^{-4}$                    & $5.87\smalltimes 10^{-4}$                        & 3.00\,\%   \\ \hline
					$25\,\nm$  & $7.56\smalltimes 10^{-4}$                    & $7.35\smalltimes 10^{-4}$                        & 2.91\,\%   \\ \hline
					$50\,\nm$  & $1.51\smalltimes 10^{-3}$                    & $1.47\smalltimes 10^{-3}$                        & 3.04\,\%   \\ \hline
				\end{tabular}
			\end{center}
			\label{tab:p_vs_p}
		\end{minipage}
	}
\end{table}

\subsubsection*{Notation}
\begin{table}[H]
	{\footnotesize
		\begin{minipage}{\textwidth}
			\captionsetup{justification=raggedright,
				singlelinecheck=false
			}
			\caption{Summary of quantities used}
			\begin{center}
				\def\arraystretch{1.5}
				\begin{tabular}{c|l}
					Quantity                                                  & Description                                                                                                        \\ \hline\hline
					$\phi$                                                    & (Integral) fluence (defined in Eq. \ref{eq:fluence})                                                               \\ \hline
					$\frac{\dd\phi}{\dd E}$                                   & Spectral fluence                                                                                                   \\ \hline
					$E\,\frac{\dd\phi}{\dd E}$                                & Spectral energy fluence                                                                                            \\ \hline
					$\phi_1^s$                                                & Fluence of particle of type $s$ produce by first simulation                                                        \\ \hline
					$\phi_2^{s\to s^\prime}$                                  & Fluence of particles of type $s^\prime$ produced by a particle of type $s$ entering the NP simulation              \\ \hline
					$\bar n^\gamma(\rnp, \phi) / \phi$                        & Mean number of photon interactions per fluence for a given fluence $\phi$ (defined in Eq.~\ref{eq:n-bar_per_phi})  \\ \hline $N_{F_k}$                                                                & Number of ionization clusters of size \mbox{$\nu\ge k$}                                                            \\ \hline
					\raisebox{-2ex}{$(N_{F_k})_\mathrm{net}^{e\to e}$}        & Net gold contribution to the number of ionization clusters of size \mbox{$\nu\ge k$} considering only those        \\ [-2ex]
					                                                          & created by electrons created by electrons (defined in Eq.\ref{eq:background-correction})                           \\ \hline
					$\big<p_i^\gamma \big>_{\phi_1^\gamma}$                   & Probability of a photon ionization within a gold NP (defined in Eq.~\ref{eq:p_g})                                  \\ \hline
					\raisebox{-2ex}{$\left(N_{F_k}\right)_\mathrm{C}^\gamma$} & Net gold contribution to the number of ionization clusters of size \mbox{$\nu\ge k$} conditional on a photon       \\ [-2ex]
					                                                          & interaction occurring within the gold NP (defined in Eq.~\ref{eq:dQdx_C})                                          \\ \hline
					$\varepsilon$                                             & Energy imparted                                                                                                    \\ \hline
					$D$                                                       & Absorbed dose, defined as energy imparted per mass $D = \frac{\dd \varepsilon}{\dd m}$                             \\ \hline
					\raisebox{-2ex}{$\mathrm{DEF}$}                           & Dose enhancement factor, the ratio of absorbed dose in a volume surrounding a gold NP to the absorbed              \\ [-2ex]
					                                                          & dose in the absence of the gold NP                                                                                 \\ \hline
					\raisebox{-2ex}{$\mathrm{CEF}_{F_k}$}                     & Clustering enhancement factor, defined as the ratio of frequency of clusters of size \mbox{$\nu\ge k$} in a volume \\ [-2ex]
					                                                          & surrounding a gold NP to the frequency of such clusters in the absence of the gold NP                              \\ \hline
				\end{tabular}
			\end{center}
			\label{tab:quantities}
		\end{minipage}
	}
\end{table}

\subsubsection*{Fluence spectra}

\begin{figure}[H]
	\fbox{
		\begin{minipage}{\textwidth}
			\begin{minipage}{.5\textwidth}
				\subfloat[]{\raggedright{}\label{fig:surface_fluence_gg}}\vspace{-0.35cm}
				\begin{center}
					\includegraphics[width=\textwidth]{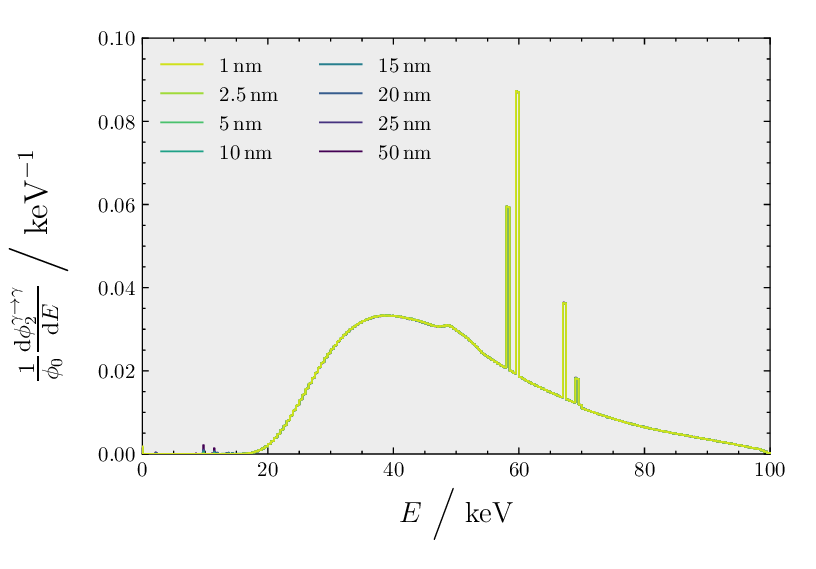}
				\end{center}
			\end{minipage}%
			\begin{minipage}{.5\textwidth}
				\subfloat[]{\raggedright{}\label{fig:surface_fluence_gg_diff}}\vspace{-0.35cm}
				\begin{center}
					\includegraphics[width=\textwidth]{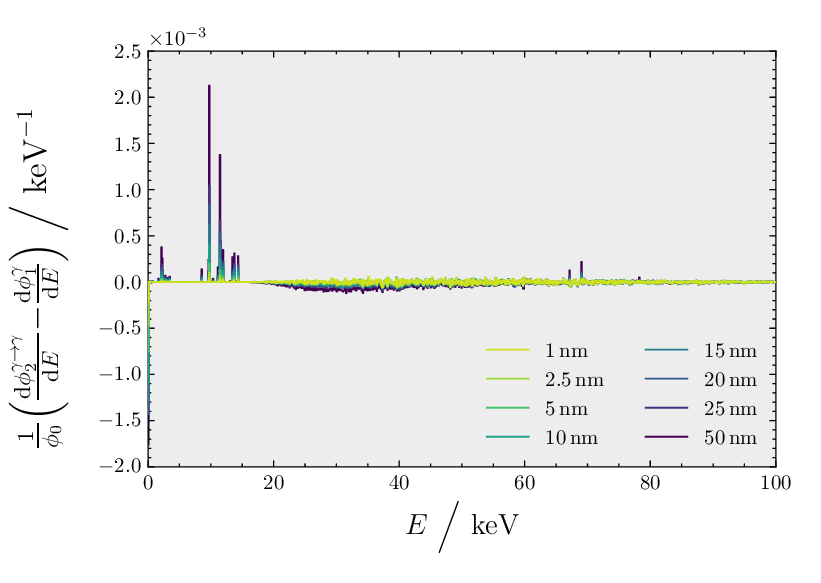}
				\end{center}
			\end{minipage}
			\caption{
				\textbf{\protect\subref{fig:surface_fluence_gg}} Spectral fluence of
				photons leaving the nanoparticle from photons entering the
				nanoparticle normalized to primary fluence for the different
				nanoparticle radii and
				\textbf{\protect\subref{fig:surface_fluence_gg_diff}}
				difference between spectral fluences of photons emitted photons
				leaving the nanoparticle and the photons entering it, normalized to the
				primary fluence for different nanoparticle radii.
			}
			\label{fig:surface_fluence_gg_both}
		\end{minipage}
	}
\end{figure}

\begin{figure}[H]
	\fbox{
		\begin{minipage}{\textwidth - 4mm}
			\begin{minipage}{.5\textwidth}
				\subfloat[]{\raggedright{}\label{fig:fluence-ge-25-comparison}}\vspace{-0.35cm}
				\begin{center}
					\includegraphics[width=\textwidth]{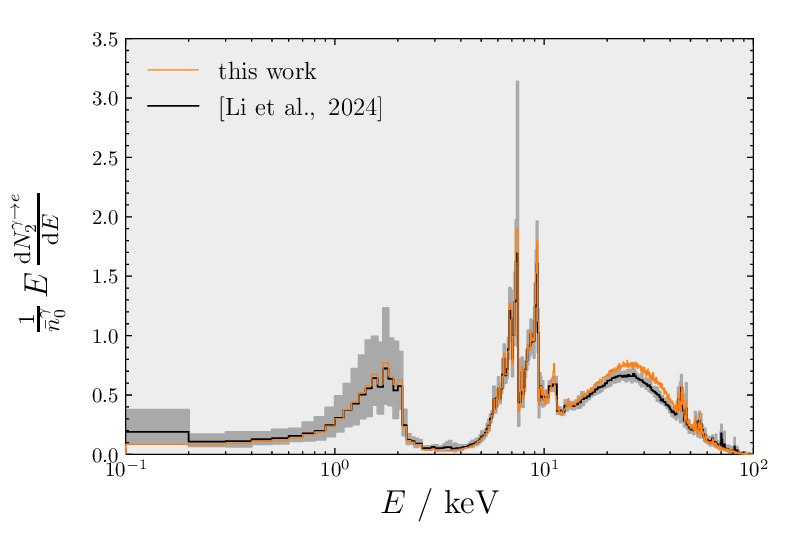}
				\end{center}
			\end{minipage}%
			\begin{minipage}{.5\textwidth}
				\subfloat[]{\raggedright{}\label{fig:fluence-ge-50-comparison}}\vspace{-0.35cm}
				\begin{center}
					\includegraphics[width=\textwidth]{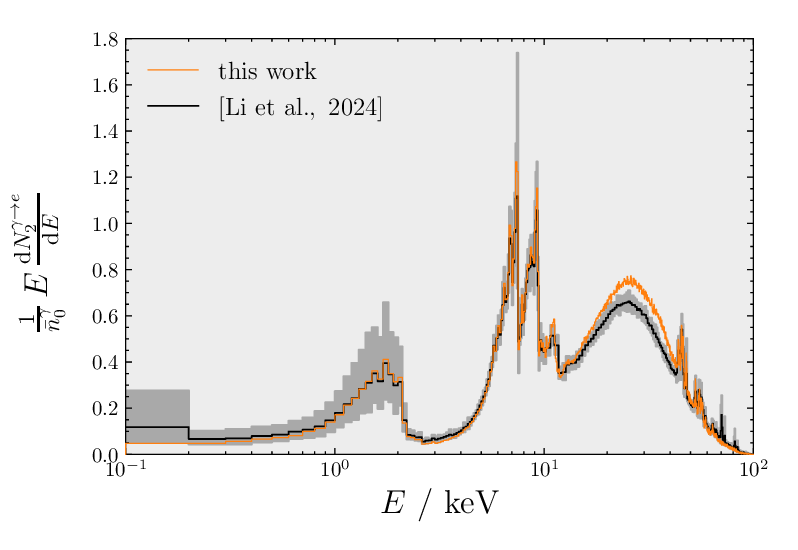}
				\end{center}
			\end{minipage}
			\begin{center}
				\begin{minipage}{.5\textwidth}
					\subfloat[]{\raggedright{}\label{fig:phi_1_per_phi_1}}\vspace{-0.35cm}
					\begin{center}
						\includegraphics[width=\textwidth]{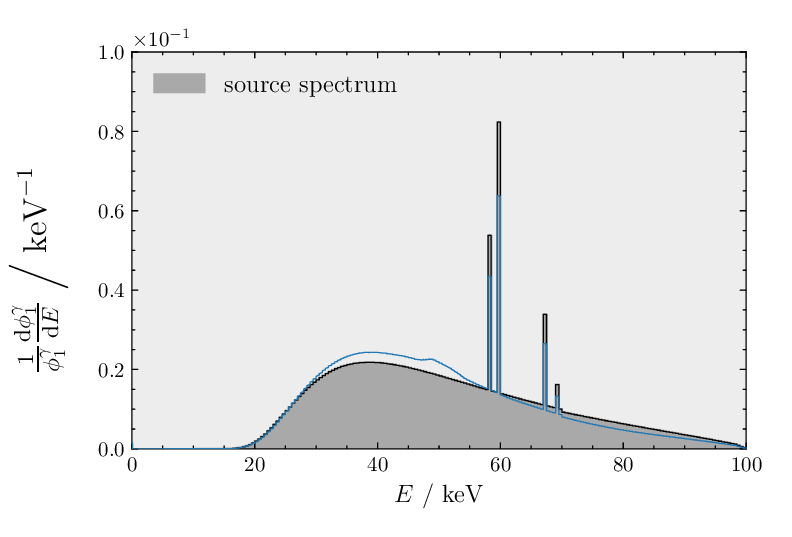}
					\end{center}
				\end{minipage}
			\end{center}

			\caption{
				Electron energy times frequency density per photon interaction
				from this work (orange line, computed from the data previously shown as
				spectral energy fluence in Fig.~\ref{fig:surface_fluence_ge}, albeit
                here with a constant bin-widths of $\mathrm{d}E = 100\,\eV$ for
				comparison) as well as work by Li et al. (2024, in press) (black line
				and gray area) for gold nanoparticles of
				\textbf{\protect\subref{fig:fluence-ge-25-comparison}}
				\mbox{$\rnp = 25\,\nm$} and
				\textbf{\protect\subref{fig:fluence-ge-50-comparison}}
				\mbox{$\rnp = 50\,\nm$}. The uncertainty bounds (gray area) correspond
				to an estimated $95\,\%$ confidence interval.
				The data are in good agreement; a slight increase in electrons of
				energies $\gtrsim 10\,\keV$ may be attributed to a larger fraction of
				Compton photons scattered back into the scoring volume. As outlined,
				this is affected by geometry, which for one affects the fraction
				$\phi_1^\gamma / \phi_0$, but also affects the spectral fluence
				qualitatively, as it increases the number of lower-energy photons.
				This can be seen in \textbf{\protect\subref{fig:phi_1_per_phi_1}},
				the spectral fluence of the primary photon spectrum, normalized to total primary fluence,
				as in Fig.~\protect\subref*{fig:fluence_prior_g} (gray shaded), as well as the
				resulting photon field in the scoring volume (blue line), as the one reported in
				Fig.~\protect\subref*{fig:fluence_prior_g}, albeit not normalized to the primary
				fluence but to $\phi_1^\gamma$.
			}
			\label{fig:fluence-ge-comparison}
		\end{minipage}
	}
\end{figure}

\begin{table}[H]
	{\footnotesize
		\begin{minipage}{\textwidth}
			\captionsetup{justification=raggedright,
				singlelinecheck=false
			}
			\caption{Ratios of the integral fluence of emitted electrons produced
				by photons to that of electrons produced by electrons for different
				gold NP radii and the same primary fluence for different energy
				ranges.
				While the ratio increases proportional to the NP radius for
				photoelectrons ($E\ge11.5$) and almost proportional for to radius for
				L shell Auger electrons the ($5.5\,\keV \le E <
					11.5\,\keV$), the increase at energies corresponding to M shell 
                    Auger electrons ($500\,\eV \le E < 5.5\,\keV$) and at energies
                    below $500\,\eV$ shows a sub-linear dependence on NP radius
			}
			\begin{center}
				\def\arraystretch{1.5}
				\begin{tabular}{l|c|c|c|c|c}
					\multirow{2}{*}{$\rnp$} & \multicolumn{5}{c}{$\phi_2^{\gamma\to e}/\phi_2^{e\to e}$}                                                                                               \\ \cline{2-6}
					                        & $E<500\,\eV$                                               & $500\,\eV \le E < 5.5\,\keV$ & $5.5\,\keV \le E < 11.5\,\keV$ & $ E \ge 11.5\,\keV$ & all   \\ \hline
					$1\,\nm$                & 6.36                                                       & 2.90                         & 0.77                           & 0.21                & 1.43  \\ \hline
					$2.5\,\nm$              & 8.76                                                       & 6.93                         & 1.91                           & 0.52                & 2.44  \\ \hline
					$5\,\nm$                & 10.72                                                      & 12.72                        & 3.81                           & 1.05                & 3.79  \\ \hline
					$10\,\nm$               & 13.53                                                      & 20.11                        & 7.69                           & 2.10                & 5.99  \\ \hline
					$15\,\nm$               & 15.02                                                      & 23.48                        & 11.57                          & 3.14                & 7.63  \\ \hline
					$20\,\nm$               & 16.19                                                      & 25.59                        & 15.48                          & 4.21                & 9.13  \\ \hline
					$25\,\nm$               & 17.09                                                      & 27.46                        & 19.29                          & 5.28                & 10.55 \\ \hline
					$50\,\nm$               & 20.32                                                      & 34.98                        & 35.87                          & 10.52               & 16.94 \\ \hline
				\end{tabular}
			\end{center}
			\label{tab:phi_ge_ee}
		\end{minipage}
	}
\end{table}

\subsubsection*{Enhancement Factors}
\begin{figure}[H]
	\fbox{
		\begin{minipage}{\textwidth}
			\begin{minipage}{.5\textwidth}
				\subfloat[]{\raggedright{}\label{fig:DEF-25-comparison}}\vspace{-0.35cm}
				\begin{center}
					\includegraphics[width=\textwidth]{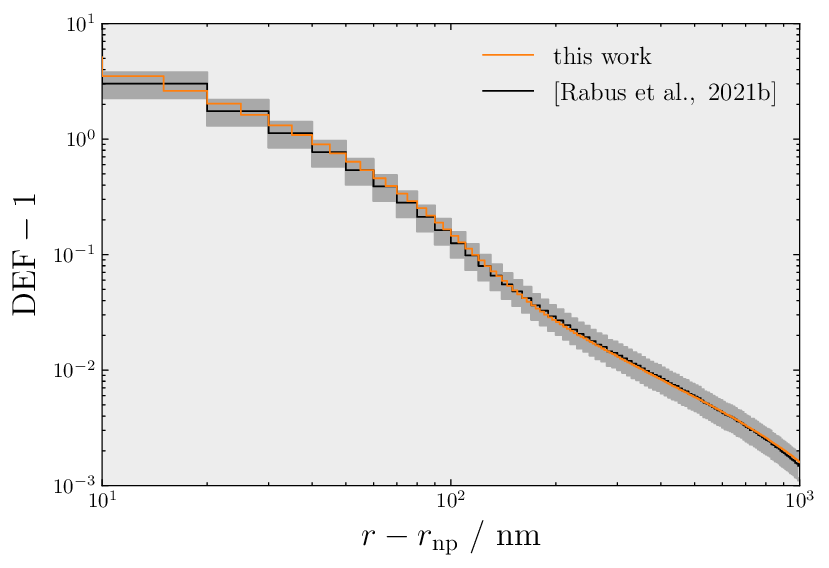}
				\end{center}
			\end{minipage}%
			\begin{minipage}{.5\textwidth}
				\subfloat[]{\raggedright{}\label{fig:DEF-50-comparison}}\vspace{-0.35cm}
				\begin{center}
					\includegraphics[width=\textwidth]{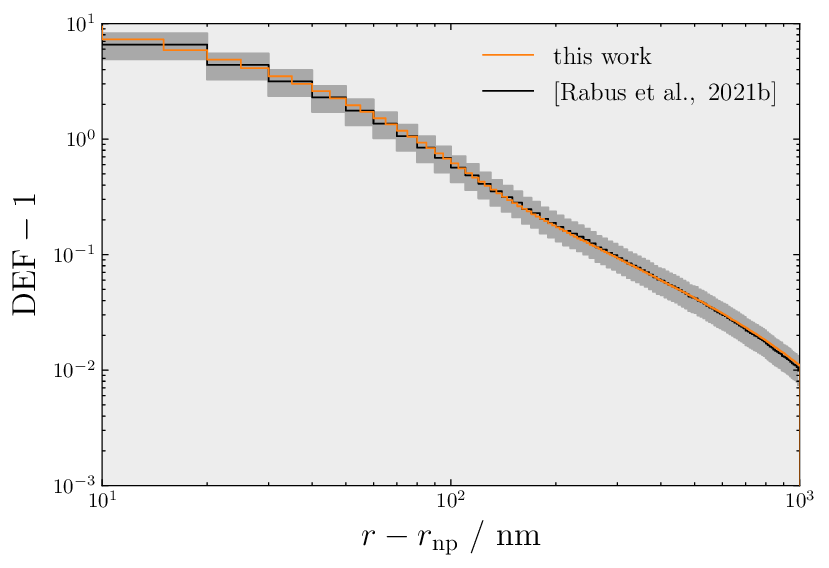}
				\end{center}
			\end{minipage}
			\caption{
				Comparison of the deviation of the dose enhancement factors from the one
				computed in this work (orange line) as well as from the EURADOS
				intercomparison, Fig.~6(b) and Fig.~6(d) in \citep{Rabus_Li_2021}
				(black line and gray area) for nanoparticles of
				\textbf{\protect\subref{fig:DEF-25-comparison}}
				\mbox{$\rnp = 25\,\nm$} and
				\textbf{\protect\subref{fig:DEF-50-comparison}}
				\mbox{$\rnp = 50\,\nm$}.
				The black line corresponds to the mean value amongst the reported
				results (excluding submitted results failing the applied plausibility
				checks).
				The reported uncertainties correspond to an estimated $95\,\%$
				confidence interval considering the scatter of these data as well as
				uncertainties associated with the procedure applied of correcting the
				lack of secondary particle equilibrium in the simulation.
				Note: The reported values in \citep{Rabus_Li_2021} correspond to upper
				bin edges. In order to maintain consistency throughout this work, the
				data has been plotted as steps, where each horizontal step represents
				the bin. While the underlying data remains identical, they visually
				differ slightly from the original work.
			}
			\label{fig:DEF-comparison}
		\end{minipage}
	}
\end{figure}

\begin{figure}[H]
	\fbox{
		\begin{minipage}{\textwidth}
			\begin{minipage}{.5\textwidth}
				\subfloat[]{\raggedright{}\label{fig:CEF}}\vspace{-0.35cm}
				\begin{center}
					\includegraphics[width=\textwidth]{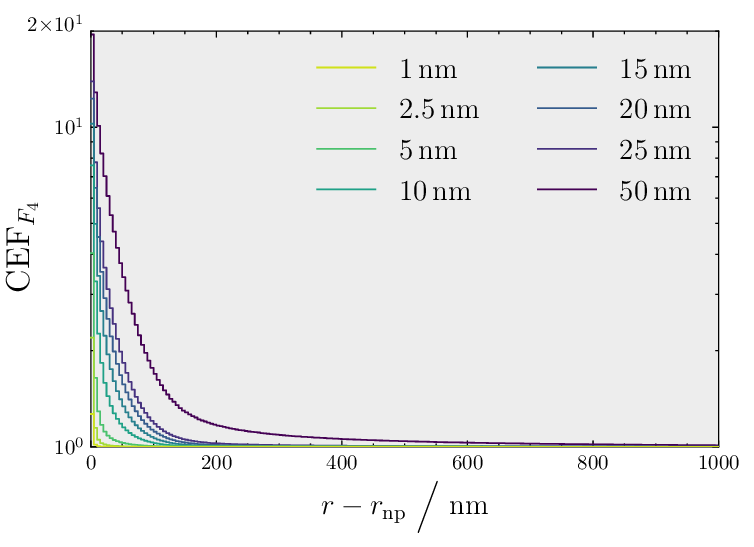}
				\end{center}
			\end{minipage}%
			\begin{minipage}{.5\textwidth}
				\subfloat[]{\raggedright{}\label{fig:DEF}}\vspace{-0.35cm}
				\begin{center}
					\includegraphics[width=\textwidth]{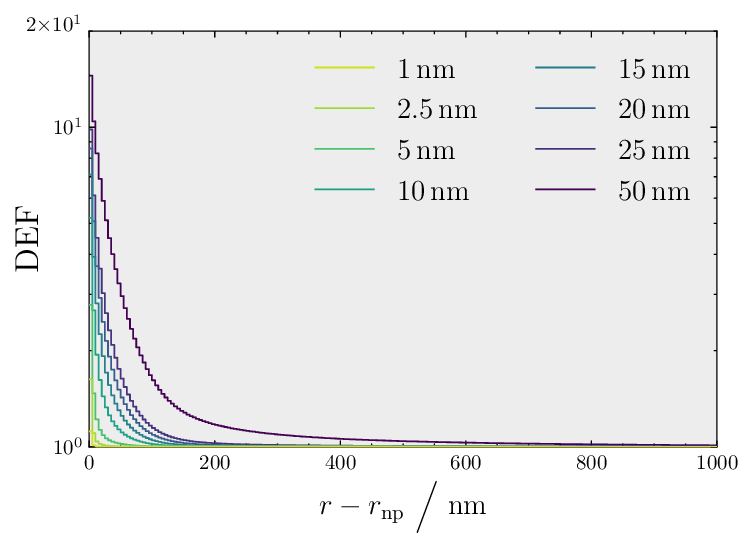}
				\end{center}
			\end{minipage}
			\caption{
				\textbf{\protect\subref{fig:CEF}} Enhancement factor of clusters of size
				\mbox{$\nu\ge 4$}, defined analogously to the DEF as the ratio of cluster
				frequency compared to the cluster frequency in the absence of the gold NP.
				\textbf{\protect\subref{fig:DEF}} Dose enhancement factor (DEF), defined as the
				ratio of absorbed dose surrounding a NP (here in a sphere) to the dose in
				the absence of the gold NP.
			}
			\label{fig:enhancement_ratios}
		\end{minipage}
	}
\end{figure}

\end{document}